\documentclass[fleqn,usenatbib]{mnras}

\usepackage{newtxtext,newtxmath}

\usepackage[T1]{fontenc}

\DeclareRobustCommand{\VAN}[3]{#2}
\let\VANthebibliography\thebibliography
\def\thebibliography{\DeclareRobustCommand{\VAN}[3]{##3}\VANthebibliography}

\usepackage{graphicx}	
\usepackage{amsmath}	
\usepackage{multirow}   
\usepackage[utf8]{inputenc} 
\usepackage{tabularx}
\usepackage{comment}
\usepackage{pdflscape}
\usepackage{scalerel}
\usepackage{orcidlink}
\usepackage{stfloats}

\usepackage{soul}

\usepackage{subcaption}

\newcommand{\plotsidesize}[2]
 {\centering \leavevmode \includegraphics[width={#2\textwidth}]{#1}}

\newcommand{\GIZMO}{{\small GIZMO}}
\newcommand{\Hmol}{{\rm H}_{2}}

\newcommand{\nH}{n_{\rm H}}

\newcommand{\cmcubed}{ {\rm cm}^{-3} }
\newcommand{\Zcrit}{Z_{\rm crit}}
\newcommand{\zfinal}{z_{\rm final}}

\newcommand{\Zsol}{{\rm Z}_{\sun}}
\newcommand{\Msol}{{M}_{\sun}}
\newcommand{\Rvir}{R_{\rm vir}}
\newcommand{\Mvir}{M_{\rm vir}}
\newcommand{\Mdust}{M_{\rm dust}}
\newcommand{\Mstar}{M_{*}}
\newcommand{\OH}{12+{\rm log_{10}(O/H)}}
\newcommand{\Tdust}{T_{\rm dust}}
\newcommand{\teq}{\tau_{\rm equil}}

\defcitealias{choban_2022:GalacticDustmodellingDust}{C22}
\defcitealias{choban_2024:DustyLocaleEvolution}{C24}

\newcommand{\gizmourl}{\href{http://www.tapir.caltech.edu/~phopkins/Site/GIZMO.html}{\url{http://www.tapir.caltech.edu/~phopkins/Site/GIZMO.html}}}

\newcommand{\datastatement}[1]{\begin{small}\section*{Data Availability Statement}\end{small}{\noindent #1}\vspace{5pt}}

\title[Dust Buildup at $z \gtrsim 5$]{A Dusty Dawn: Galactic Dust Buildup at $z \gtrsim 5$}
\author[C. R. Choban et al.]{
\parbox[t]{\textwidth}{
        Caleb R. Choban\orcidlink{0000-0001-9200-169X}$^{1}$\thanks{email: cchoban@iu.edu},
        Samir Salim\orcidlink{0000-0003-2342-7501}$^{1}$,
        Du\v{s}an Kere\v{s}\orcidlink{0000-0002-1666-7067}$^{2,3}$,
        Christopher C. Hayward\orcidlink{0000-0003-4073-3236}$^{4}$,
        and Karin M. Sandstrom\orcidlink{0000-0002-4378-8534}$^{2}$
} \vspace*{4pt} \\
$^{1}$ Department of Astronomy, Indiana University, Bloomington, IN 47405, USA \\
$^{2}$ Department of Astronomy \& Astrophysics, University of California at San Diego, La Jolla, CA 92093, USA \\
$^{3}$ Department of Physics, University of California at San Diego, La Jolla, CA 92093, USA \\
$^{4}$ Center for Computational Astrophysics, Flatiron Institute, New York, NY 10010 USA \\
}

\date{Accepted XXX. Received YYY; in original form ZZZ}

\pubyear{2024}

\begin{document}
\label{firstpage}
\pagerange{\pageref{firstpage}--\pageref{lastpage}}

\maketitle

\begin{abstract}
Over the last decade, the Atacama Large Millimeter Array has revealed massive, dusty star-forming galaxies at $z\gtrsim5$, and the James Webb Space Telescope is primed to uncover even more information about them. 
These observations need dust evolution theory to provide context and are excellent benchmarks to test this theory.
Here, we investigate the evolution of galactic dust budget at cosmic dawn using a suite of cosmological zoom-in simulations of moderately massive, high-redshift ($\Mstar\gtrsim10^9\Msol$; $z\gtrsim5$) galaxies from the FIRE project, the highest resolution ($m_{\rm b} \approx 7100\,\Msol$) of such simulations to date.
Our simulations incorporate a dust evolution model that accounts for the dominant sources of dust production, growth, and destruction and follows the evolution of specific dust species, allowing it to replicate a wide range of present-day observations.
We find, similar to other theoretical works, that dust growth via gas-dust accretion is the dominant producer of dust mass for these massive, $z\gtrsim 5$ galaxies. 
However, our fiducial model produces $\Mdust$ that fall ${\gtrsim}1$ dex below observations at any given $\Mstar$ (typical uncertainties are ${\sim}1$ dex), which we attribute to reduced accretion efficiencies caused by a combination of low galactic metallicities and extremely bursty star formation.
Modest enhancements (i.e., within observational/theoretical uncertainties) to accretion and SNe II dust creation raise $\Mdust$ by ${\lesssim}1$ dex, but this still falls below observations which assume $\Tdust\sim25$ K. 
One possibility is that inferred dust masses for $z\gtrsim4$ galaxies are overestimated, and recent observational/analytical works that find $\Tdust\sim50$ K along with metallicity constraints tentatively support this.

\end{abstract}




\section{Introduction}

Until recently, our understanding of dust and its abundance in the early universe was `shrouded' in mystery.
The advent of the Atacama Large Millimeter Array (ALMA) in the last decade has parted the proverbial curtain on rest-frame far-infrared (FIR) light at this epoch, allowing for unprecedented detection of dust continuum sources at $z>4$. 
In the succeeding period, numerous massive, extremely dusty star-forming galaxies (DSFGs) have been discovered at $z\gtrsim5$, starting with individual detections 
\citep{vieira_2013:DustyStarburstGalaxies,hezaveh_2013:ALMAObservationsSPTdiscovered,watson_2015:DustyNormalGalaxy,spilker_2016:ALMAImagingGravitational,laporte_2017:DustReionizationEra,strandet_2017:ISMPropertiesMassive,miller_2018:MassiveCoreCluster,tamura_2019:DetectionFarinfraredIii,reuter_2020:CompleteRedshiftDistribution,witstok_2023:EmpiricalStudyDust} and recently expanding to ${\sim}100$ galaxies with the ALMA Large Program to INvestigate [\textsc{C ii}] at Early Times \citep[ALPINE;][]{lefevre_2020:ALPINEALMACIISurvey, bethermin_2020:ALPINEALMACIISurvey, faisst_2020:ALPINEALMAIISurvey} $z\sim5$ survey, the Reionization Era Bright Emission Line Survey \citep[REBELS;][]{bouwens_2022:ReionizationEraBright} $z\sim7$ survey, and the Systematic Exploration in the Reionization Epoch using Nebular And Dust Emission \citep[SERENADE;][]{mitsuhashi_2024:SERENADEIIALMA} $z\sim6$ survey. Furthermore, the dust in one $z\sim5.7$ galaxy has even been spatially-resolved as part of the [\textsc{C ii}] Resolved ISM in STar-forming
galaxies with ALMA \citep[CRISTAL;][]{villanueva_2024:ALMACRISTALSurveyDust} survey.
However, these samples are primarily UV-selected\footnote{ALMA has also observed galaxies detected in the submillimetre by the South Pole Telescope~\citep{vieira_2013:DustyStarburstGalaxies,hezaveh_2013:ALMAObservationsSPTdiscovered,spilker_2016:ALMAImagingGravitational,miller_2018:MassiveCoreCluster,reuter_2020:CompleteRedshiftDistribution}, 
finding they are generally the most massive DSFGs (median SFR $\sim10^3 \Msol/{\rm yr}$ and $\Mdust\sim10^9\Msol$). However, these observations have no accompanying stellar mass estimates. There have also been a few serendipitous detections of $z>4$ IR-bright galaxies with no accompanying UV detection \citep{gruppioni_2020:ALPINEALMACIISurvey,fudamoto_2021:NormalDustobscuredGalaxies}, with measured $\Mstar\lesssim10^{10}\Mstar$ falling within the range of the ALPINE/REBELS sample.}
and are, therefore, biased towards a subset of UV-bright galaxies. Furthermore, the information these surveys provide is primarily restricted to estimates of galactic dust and stellar mass. These limitations make determining the exact process responsible for such dusty galaxies difficult.
In particular, both observations \citep{remy-ruyer_2014:GasdustMassRatios,devis_2019:SystematicMetallicityStudy} and simulations, which we discuss later, find the dust content of local galaxies has a strong correlation with galactic metallicity.
Therefore, any predictions of dust evolution in these high-z galaxies are largely predicated on their assumed chemical evolution \citep[e.g.][]{palla_2020:InfluenceTopheavyIntegrated,palla_2024:MetalDustEvolution}. 

Fortuitously, the recently launched James Webb Space Telescope (JWST) probes the rest-frame optical at $z>4$, opening another observational window for high-z DSFGs. 
Notably, JWST can resolve prominent nebular emission lines, providing measurements of galactic gas-phase metallicity. 
This has already led to estimates of the gas-phase mass-metallicity relation, which suggests high-z galaxies are more metal-poor than their present-day stellar mass counterparts \citep{nakajima_2023:JWSTCensusMassMetallicity,curti_2024:JADESInsightsLowmass,chemerynska_2024:ExtremeLowmassEnd}.
However, there is a large scatter in this relation\footnote{There is also a large ${\gtrsim}1$ dex scatter in the metallicity of damped Lyman-$\alpha$ systems probed out to $z\sim5$ \citep{wiseman_2017:EvolutionDustmetalsRatio,peroux_2020:CosmicBaryonMetal}.}, and some observations of individual galaxies at $z{\sim}4$ find near solar metallicities \citep{birkin_2023:JWSTsTEMPLATESStar}.
Furthermore, JWST can resolve attenuation curves, revealing information on dust population composition, such as the presence of the 2175 \r{A} bump produced by small carbonaceous grains \citep{markov_2023:DustAttenuationLaw,witstok_2023:CarbonaceousDustGrains}. JWST can also identify extremely red galaxies that are heavily dust-obscured at rest-frame UV wavelengths \citep{akins_2023:TwoMassiveCompact}, avoiding the UV-selection bias inherent in many previous observations.

Given current and forthcoming DSFGs observations, dust evolution theory needs to be taken to task to provide context for observations, further our understanding of the dust life cycle, and make subsequent predictions that can be followed up with JWST.
Indeed, not long after the first detections of high-z DSFGs, analytical models showed that the creation of dust by supernovae (SNe) alone\footnote{Asymptotic giant branch (AGB) dust production is generally believed to be subdominant at these times, but this may not be the case if they reach a high enough metallicity \citep[e.g.][]{schneider_2024:FormationCosmicEvolution}.} cannot produce the dust masses observed due to the subsequent destruction of dust by supernovae shocks \citep[e.g.][]{michalowski_2015:DustProduction680850,lesniewska_2019:DustProductionScenarios}.
In the local universe, there is ample evidence that preexisting dust grains can grow from the accretion of gas-phase metals (gas-dust accretion)\footnote{These findings are not unanimous among all works. 
In particular, \citet{priestley_2022:ImpactMetallicitydependentDust} suggests the contribution of dust growth via accretion may be overestimated if both high stardust creation efficiencies and increased SNe dust destruction in low-metallicity environments are assumed and \citet{ferrara_2016:ProblematicGrowthDust} suggest gas-dust accretion is hampered in high-z environments.}. 
Notably, observations of the Milky Way (MW) and nearby galaxies find the fraction of metals locked in dust (dust-to-metals ratio; D/Z) increases with local gas surface density~\citep{jenkins_2009:UnifiedRepresentationGasPhase,roman-duval_2021:METALMetalEvolution,roman-duval_2014:DustGasMagellanic,roman-duval_2017:DustAbundanceVariations,chiang_2018:SpatiallyResolvedDustmetals,clark_2023:QuestMissingDust}. 
Given this, gas-dust accretion is generally believed to be responsible for the large dust masses seen in high-z DSFGs, but our understanding of this process is limited due to the inherent difficulties of experimental study.

In recent years, dust evolution models integrated into semi-analytical and cosmological simulations have been utilized to further our understanding of gas-dust accretion and all other processes in the dust life cycle \citep[e.g.][]{bekki_2015:CosmicEvolutionDust,mckinnon_2016:DustFormationMilky,li_2019:DustgasDustmetalRatio,granato_2021:DustEvolutionZoomcosmological,choban_2024:DustyLocaleEvolution}. Despite variations in methodologies and included physics these models agree in broad strokes, finding that dust growth via accretion is responsible for the bulk of dust content of the MW and local galaxies \citep[e.g.][]{mckee_1989:DustDestructionInterstellar,draine_2009:InterstellarDustModels}. However, details in the dust life cycle important for the predictions of high-z DSFGs vary considerably between works. This is due to a focus on present-day observations, which are the result of dust life cycle processes integrated over ${\sim}13$ Gyrs of evolutionary history.
In particular, the assumed efficiency of dust creation by SNe II varies considerably between models since it is found to have little effect on present-day observations due to the dominance of accretion.
Furthermore, while simulations predict that accretion becomes efficient (more dust is grown than is destroyed) within a galaxy once that galaxy's mass-weighted median metallicity passes a certain threshold (critical metallicity threshold; $\Zcrit$) \citep{inoue_2011:OriginDustGalaxies,asano_2013:DustFormationHistory,zhukovska_2014:DustOriginLatetype,feldmann_2015:EquilibriumViewDust,popping_2017:DustContentGalaxies,hou_2019:DustScalingRelations,li_2019:DustgasDustmetalRatio,graziani_2020:AssemblyDustyGalaxies,triani_2020:OriginDustGalaxies,parente_2022:DustEvolutionMUPPI,choban_2024:DustyLocaleEvolution}, the predicted thresholds vary considerably ($\Zcrit \sim 0.03-0.5 \,\Zsol$). Works that track the evolution of chemically distinct dust species further suggest a separate $\Zcrit$ for each dust species \citep{granato_2021:DustEvolutionZoomcosmological,parente_2022:DustEvolutionMUPPI,choban_2024:DustyLocaleEvolution}. 
There is also a general consensus that accretion steadily increases the galactic D/Z, and dust mass, up to an equilibrium value once $\Zcrit$ is reached. However, the exact timescale of this buildup varies between galaxies, depending on the ISM phase structure, star formation rate (SFR), and the initial amount of dust \citep[e.g.][]{zhukovska_2008:EvolutionInterstellarDust,zhukovska_2014:DustOriginLatetype}, and could be ${\gtrsim}1$ Gyr for a MW-mass galaxy \citep{choban_2024:DustyLocaleEvolution}.

Fewer works have focused on dust evolution in high-z DSFGs, utilizing semi-analytical models \citep{popping_2017:DustContentGalaxies,vijayan_2019:DetailedDustModelling,triani_2020:OriginDustGalaxies,dayal_2022:ALMAREBELSSurvey,mauerhofer_2023:DustEnrichmentEarly} and cosmological simulations \citep{graziani_2020:AssemblyDustyGalaxies,esmerian_2022:ModelingDustProduction,esmerian_2024:ModelingDustProduction,lewis_2023:DUSTiERDUSTEpoch,dicesare_2023:AssemblyDustyGalaxies,lower_2023:CosmicSandsOrigin,lower_2024:CosmicSandsII}, and their predictions vary considerably due to differing galaxy and dust model prescriptions.
While many authors agree that accretion is the dominant producer of dust mass at this epoch, others predict accretion is inefficient and that SNe II dominate dust production \citep{triani_2020:OriginDustGalaxies,dayal_2022:ALMAREBELSSurvey}. 
Furthermore, some models struggle to reproduce the high dust masses observed for high-z DSFGs \citep{vijayan_2019:DetailedDustModelling,triani_2020:OriginDustGalaxies,dayal_2022:ALMAREBELSSurvey}, while others reproduce them with relative ease \citep{popping_2017:DustContentGalaxies,dicesare_2023:AssemblyDustyGalaxies,lewis_2023:DUSTiERDUSTEpoch,lower_2023:CosmicSandsOrigin}.
A major limitation of the above-mentioned simulations is their inability to resolve the multi-phase ISM and self-consistently model dust life cycle processes. 
In particular, they utilize `tuned' sub-resolution prescriptions for star formation, affecting the predicted galactic star formation history \citep[e.g.][]{iyer_2020:DiversityVariabilityStar}. They also use sub-resolution gas-dust accretion routines and do not track the evolution of chemically distinct dust species, affecting the predicted dust population evolution and spatial variability within galaxies \citep{choban_2022:GalacticDustmodellingDust,choban_2024:DustyLocaleEvolution}.

There has also been a growing trend of post-processing high-z simulations originally run without an explicit dust evolution model by using radiative transfer codes and assumed dust populations to make observational predictions for ALMA and JWST \citep{liang_2018:SubmillimetreFluxProbe,liang_2019:DustTemperaturesHighredshift,cochrane_2019:PredictionsSpatialDistribution,cochrane_2022:DustTemperatureUncertainties,cochrane_2023:PredictingSubmillimetreFlux,cochrane_2024:DisappearingGalaxiesOrientation,ma_2019:DustAttenuationDust,vogelsberger_2020:HighredshiftJWSTPredictions,shen_2020:HighredshiftJWSTPredictions,parsotan_2021:RealisticMockObservations,pallottini_2022:SurveyHighzGalaxies,shen_2022:HighredshiftPredictionsIllustrisTNG,vijayan_2022:FirstLightReionisation,katz_2023:SPHINXPublicData}.
The dust populations assumed are typically derived from MW observations and are relatively simplistic (e.g. D/Z $=0.4$ for gas with $T\lesssim10^6$ K). However, high-z dust populations can be drastically different from the MW, affecting predicted observables \citep{lower_2024:CosmicSandsII}, and so accurate dust population predictions from high-z simulations are needed.

\begin{table*}
        \renewcommand{\arraystretch}{1.15}
	\centering
	\begin{tabular}{llcccccccccc} 
		\hline
		Name & $\zfinal$ & $\Mvir$ & $\Rvir$ & $M_{*}$ & $R_{\rm *, 1/2}$ & $M_{\rm gas, neutral}$ & $R_{\rm neutral, 1/2}$ & Z & $M_{\rm dust}$ & $R_{\rm dust, 1/2}$ & SFR$_{\rm 10\,Myr}$ \\
             & & $(\Msol)$ & (kpc) & $(\Msol)$ & (kpc) & $(\Msol)$ & (kpc) & $(\Zsol)$ & $(\Msol)$ & (kpc) & $(\Msol/{\rm yr})$ \\
		\hline
            \multicolumn{12}{c}{Fiducial} \\
            \hline
            z5m11d & 5.0 & $1.28 \times 10^{11}$ & 26.07 &  $1.21 \times 10^{9}$ & 1.85 & $4.03 \times 10^{9}$ & 2.94 & 0.07 &  $1.04 \times 10^{5}$ & 2.21 & 8.07 \\
            z5m12a & 5.0 & $4.39 \times 10^{11}$ & 39.06 &  $5.43 \times 10^{9}$ & 6.23 & $1.91 \times 10^{10}$ & 5.43 & 0.12 &  $1.08 \times 10^{6}$ & 6.19 & 5.15 \\
            z5m12b & 5.1 & $7.83 \times 10^{11}$ & 47.12 &  $1.61 \times 10^{10}$ & 1.72 & $1.79 \times 10^{10}$ & 6.40 & 0.14 &  $3.26 \times 10^{6}$ & 5.50 & 3.53 \\
            z5m12d & 5.0 & $5.05 \times 10^{11}$ & 40.93 &  $8.19 \times 10^{9}$ & 6.25 & $1.98 \times 10^{10}$ & 4.33 & 0.16 &  $1.29 \times 10^{6}$ & 3.76 & 5.00 \\
            z7m12a & 8.1 & $2.63 \times 10^{11}$ & 21.80 &  $2.09 \times 10^{9}$ & 1.84 & $2.06 \times 10^{10}$ & 2.35 & 0.05 &  $2.50 \times 10^{5}$ & 2.14 & 20.93 \\
            z7m12b & 7.7 & $3.40 \times 10^{11}$ & 24.83 &  $6.81 \times 10^{9}$ & 2.29 & $1.40 \times 10^{10}$ & 2.99 & 0.18 &  $9.52 \times 10^{5}$ & 2.87 & 13.00 \\
            z7m12c & 7.0 & $4.41 \times 10^{11}$ & 29.53 &  $7.77 \times 10^{9}$ & 3.37 & $1.96 \times 10^{10}$ & 2.78 & 0.10 &  $6.30 \times 10^{5}$ & 2.09 & 12.61 \\
            z9m12a & 10.8 & $2.11 \times 10^{11}$ & 15.63 &  $8.65 \times 10^{9}$ & 0.21 & $6.82 \times 10^{9}$ & 1.36 & 0.17 &  $1.58 \times 10^{6}$ & 1.36 & 263.75 \\
		\hline
            \multicolumn{12}{c}{Enhanced Stardust \& Accretion} \\
            \hline
            z5m11d\_enh & 5.0 & $1.28 \times 10^{11}$ & 25.93 &  $1.22 \times 10^{9}$ & 1.57 & $4.46 \times 10^{9}$ & 2.97 & 0.08 &  $6.81 \times 10^{5}$ & 2.61 & 0.72 \\
            z5m12a\_enh & 5.0 & $4.41 \times 10^{11}$ & 39.13 &  $4.92 \times 10^{9}$ & 5.37 & $2.27 \times 10^{10}$ & 5.11 & 0.13 &  $7.08 \times 10^{6}$ & 4.79 & 7.60 \\
            z5m12b\_enh & 5.6 & $5.65 \times 10^{11}$ & 38.64 &  $1.68 \times 10^{10}$ & 1.19 & $9.83 \times 10^{9}$ & 3.60 & 0.51 &  $2.11 \times 10^{7}$ & 4.25 & 35.09 \\
            z5m12d\_enh & 5.0 & $5.29 \times 10^{11}$ & 41.57 &  $1.10 \times 10^{10}$ & 6.89 & $2.41 \times 10^{10}$ & 5.24 & 0.17 &  $9.90 \times 10^{6}$ & 5.19 & 21.43 \\
            z7m12a\_enh & 8.0 & $2.88 \times 10^{11}$ & 22.79 &  $3.44 \times 10^{9}$ & 1.54 & $2.28 \times 10^{10}$ & 2.26 & 0.06 &  $3.29 \times 10^{6}$ & 1.83 & 77.03 \\
            z7m12b\_enh & 8.0 & $3.16 \times 10^{11}$ & 23.49 &  $9.48 \times 10^{9}$ & 1.45 & $1.22 \times 10^{10}$ & 2.29 & 0.13 &  $6.76 \times 10^{6}$ & 1.04 & 46.10 \\
            z7m12c\_enh & 7.0 & $4.50 \times 10^{11}$ & 29.72 &  $1.06 \times 10^{10}$ & 2.30 & $1.94 \times 10^{10}$ & 3.20 & 0.19 &  $1.11 \times 10^{7}$ & 2.31 & 50.22 \\
            z9m12a\_enh & 10.4 & $2.63 \times 10^{11}$ & 17.47 &  $5.93 \times 10^{9}$ & 1.09 & $8.38 \times 10^{9}$ & 2.87 & 0.16 &  $5.15 \times 10^{6}$ & 2.93 & 65.30 \\
	   \hline
	\end{tabular}
	\caption{Parameters describing properties of simulated galaxies, all run with $7100\,\Msol$ mass resolution for gas and star particles. The "Fiducial" group of simulations utilized our "Species" dust evolution model as presented in \citetalias{choban_2022:GalacticDustmodellingDust}. The "Enhanced Stardust \& Accretion" increases the SNe II dust creation efficiencies and decreases the gas-dust accretion timescale by a factor of 4 (see Table~\ref{tab:model_comparison} for quantitative comparison). 
    \textbf{(1)} Name of simulation. 
    \textbf{(2)} The redshift the simulation is run to.
    \textbf{(3)} Virial mass of dark matter halo at $\zfinal$. 
    \textbf{(4)} Virial radius of dark matter halo at $\zfinal$. 
    \textbf{(5)} Stellar mass within virial radius at $\zfinal$. 
    \textbf{(6)} The stellar half mass radius within virial radius at $\zfinal$. 
    \textbf{(7)} Neutral ($\textsc{H\,i}$ + $\Hmol$) gas mass within virial radius at $\zfinal$. 
    \textbf{(8)} Neutral gas half mass radius within $0.2\Rvir$ at $\zfinal$.     
    \textbf{(9)} Median metallicity for cool ($T<1000$ K) gas within virial radius at $\zfinal$. 
    \textbf{(10)} Dust mass within virial radius at $\zfinal$. 
    \textbf{(11)} Dust half mass radius within $0.2\Rvir$ at $\zfinal$.
    \textbf{(12)} Star formation rate averaged over the last 10 Myr within the virial radius.}
    \label{tab:simulations}
\end{table*}

In this work, we present a subset of cosmological zoom-in simulations of moderately massive, high-redshift ($\Mstar\gtrsim10^9\Msol$; $z\gtrsim5$) galaxies from the Feedback in Realistic Environments (FIRE) project\footnote{\url{http://fire.northwestern.edu}}, originally simulated in \citet{ma_2018:SimulatingGalaxiesReionization,ma_2019:DustAttenuationDust}, rerun with the integrated ``Species'' dust evolution model presented in \citet[][\citetalias{choban_2022:GalacticDustmodellingDust} hereafter]{choban_2022:GalacticDustmodellingDust}. This model is able to reproduce numerous galaxy-integrated and spatially-resolved observations of dust in the local universe due to its ability to track the evolution of specific dust species with set chemical compositions and incorporation of a physically motivated dust growth routine \citep[][\citetalias{choban_2024:DustyLocaleEvolution} hereafter]{choban_2024:DustyLocaleEvolution}.
Notably, this is the first application of such a dust evolution model in high-redshift simulations that resolve the multi-phase ISM and giant molecular clouds.
We find that accretion is the dominant producer of dust in these massive, high-z galaxies, but their extreme burstiness and low metallicity reduce the efficiency of dust mass buildup via accretion.
Compared to observations, our model produces systematically lower dust masses (${\gtrsim} 1$ dex) than inferred dust masses of $z\gtrsim5$ DSFGs.
Even with reasonable modifications (i.e. within observational and theoretical uncertainties) to accretion timescales and SNe II dust creation routines, we can reproduce only a subset of observed DSFGs. 
However, many observed dust masses lie above metal-budget upper limits from our simulations, which match the observed high-z mass-metallicity relation \citep{marszewski_2024:HighRedshiftGasPhaseMass}. 
Therefore, we argue that the dust masses of some DSFGs are likely overestimated due to uncertainties in assumed dust temperature, which is tentatively supported by recent observations/analytical works.
We also find that these galaxies are dominated by silicate dust and have a large spread in D/Z across ISM phases, both critical aspects that should be considered when post-processing high-z simulations.
In regards to future observations, we highlight the need for estimates of the dust mass-metallicity relation for low-metallicity high-z galaxies to better constrain our understanding of dust evolution at high-z.

This paper is organized as follows. In Section~\ref{sec:methods}, we provide a brief overview of our simulation sample along with the galaxy formation and dust evolution model used. In Section~\ref{sec:results}, we present the results of our simulations, focusing on the evolution of the stellar, metal, and dust population properties for each galaxy in Section~\ref{sec:galaxy_evolution} and comparing them with high-z observations in Section~\ref{sec:results_observations}. We discuss observational uncertainties and biases for high-z DSFGs in Section~\ref{sec:observations}, uncertainties in our dust model in Section~\ref{sec:dust_model}, and compare our findings with other high-z simulations in Sec.~\ref{sec:simulation_comparison}. Finally, we present our conclusions in Section~\ref{sec:conclusions}.

\section{Methodology} \label{sec:methods}

To study the evolution of dust in massive galaxies at high redshift, we reran a subset of cosmological simulations from the HiZ FIRE-2 suite presented in \citet{ma_2018:SimulatingGalaxiesReionization,ma_2019:DustAttenuationDust}, selecting galaxies from the suite with stellar masses $M_*\gtrsim10^9\Msol$ at $z=5$, 7, or 9 which matches the lowest stellar masses with measured dust masses observed at $z\gtrsim5$.
The exact details for each simulation's final redshift ($\zfinal$), resulting galactic properties at said redshift, and mass resolution are provided in Table~\ref{tab:simulations}.

\subsection{Galaxy and Dust Evolution Models}

All simulations in this work are run with the \GIZMO\ code base \citep{hopkins_2015:NewClassAccurate} in the meshless finite-mass (MFM) mode with FIRE-2 \citep{hopkins_2018:FIRE2SimulationsPhysics} model of star formation and stellar feedback. FIRE-2, an updated version of FIRE \citep{hopkins_2014:GalaxiesFIREFeedback}, incorporates multiple sources of stellar feedback, including stellar winds (O/B and AGB stars), ionizing photons, radiation pressure, and supernovae (both Types Ia and II). Gas cooling is followed for $T=10-10^{10}$ K including free-free, Compton, metal-line, molecular, fine-structure, and dust collisional processes while gas is also heated by cosmic rays, photo-electric, and photoionization heating by both local sources and an uniform but redshift dependent meta-galactic background \citep{faucher-giguere_2009:NewCalculationIonizing}, including the effect of self-shielding (note our dust evolution model is not coupled to FIRE ISM physics as discussed in Sec.~\ref{sec:model_limitations}). Star formation is only allowed in cold, molecular, and locally self-gravitating regions with  $\nH \geq 1000 \, {\rm cm}^{-3}$.

Each star particle represents a stellar population with a known mass, age, and metallicity assuming a \citet{kroupa_2002:InitialMassFunction} initial mass function (IMF) from $0.1-100\; \Msol$. The luminosity, mass loss rates, and SNe II rates of each star particle are calculated based on the {\small STARBURST99} \citep{leitherer_1999:Starburst99SynthesisModels} libraries, and SNe Ia rates following \citet{mannucci_2006:TwoPopulationsProgenitors}. Metal yields from SNe II, Ia, and AGB winds are taken from \citet{nomoto_2006:NucleosynthesisYieldsCorecollapse}, \citet{iwamoto_1999:NucleosynthesisChandrasekharMass}, and \citet{izzard_2004:NewSyntheticModel} respectively. Evolution of eleven species (H, He, C, N, O, Ne, Mg, Si, S, Ca, and Fe) is tracked for each gas cell. Sub-resolution turbulent metal diffusion is modeled as described in \citet{su_2017:FeedbackFirstSurprisingly} and \citet{escala_2018:ModellingChemicalAbundance}. FIRE-2 adopts the older \citet{anders_1989:AbundancesElementsMeteoritic} solar metal abundances with $Z\sim 0.02$, so any future mention of solar abundances refers to the Andres \& Gravesse abundances.

FIRE is ideally suited to investigate galactic dust evolution at high redshift given its success in matching a wide range of observations related to galaxies and their evolution, including the mass-metallicity relation up to $z\gtrsim 10$ \citep{ma_2016:OriginEvolutionGalaxy,feldmann_2023:FIREboxSimulatingGalaxies,bassini_2024:InflowOutflowProperties,marszewski_2024:HighRedshiftGasPhaseMass} and the Kennicutt–Schmidt star formation law \citep{hopkins_2014:GalaxiesFIREFeedback,orr_2018:WhatFIREsStar,gurvich_2020:PressureBalanceMultiphase}. This success is owed to the high resolution, star formation criteria, cooling to low temperatures, and multi-channel stellar feedback of FIRE, all of which result in a reasonable ISM phase structure and giant molecular cloud (GMC) mass function \citep{benincasa_2020:LiveFastYoung}. These also lead to the self-consistent development of galactic winds that eject large amounts of gas \citep{muratov_2015:GustyGaseousFlows, angles-alcazar_2017:CosmicBaryonCycle} and metals \citep{muratov_2017:MetalFlowsCircumgalactic, hafen_2019:OriginsCircumgalacticMedium,pandya_2021:CharacterizingMassMomentum} out of galaxies, preventing excessive star formation and leading to a plausible stellar mass-halo mass relation.

Our simulations utilize the integrated ``Species'' dust evolution model presented in \citetalias{choban_2022:GalacticDustmodellingDust}, which we refer to the reader for full details. This model includes the present-day dominant sources of dust production, tracking and differentiating between dust created from SNe Ia and II, AGB stars, and dust growth from gas-phase metal accretion in the ISM. It includes the dominant dust destruction mechanisms, accounting for dust destroyed by SNe shocks, thermal sputtering, and astration (dust destroyed during the formation of stars). These processes are modeled self-consistently in each gas cell, depending on local gas properties (temperature, density, metallicity, etc.) and nearby star particles owing to the FIRE model's in-depth treatment of the multi-phase ISM and time-resolved individual SNe events \citep{hopkins_2018:HowModelSupernovae}. Notably, we restrict gas-dust accretion to cool ($T\leq300$ K)\footnote{This is caused by the decreasing sticking efficiency of gas-phase elements onto the surface of dust grain with temperature. However, little to no experimental data exists so our model uses a simple step function at a set cutoff temperature.} 
gas and destroy dust locally around individual SNe events, allowing us to track the local variability of dust in the ISM.
We also follow the evolution of specific dust species (carbonaceous, silicates, and silicon carbide) and theoretical oxygen-bearing (O-reservoir) and nanoparticle metallic iron (Nano-iron) dust species with set chemical compositions. 
Consequently, each dust species has a key element\footnote{Here key element refers to the element for which $n/i$ has the lowest value, where $n$ is the number abundance of the element and $i$ is the number of atoms of the element in one formula unit of the dust species under consideration.} that limits individual accretion growth rates and the maximum formable amount of said dust species. 
We also incorporate sub-resolution turbulent dust diffusion, which follows the metal diffusion prescription in FIRE, and a dense molecular gas scheme. 
This scheme is critical to account for Coulomb enhancement of gas-dust accretion in atomic/diffuse molecular gas and the reduction in carbonaceous dust accretion due to the lock-up of gas-phase C into CO in dense molecular gas. 
The above-listed details enable this model to match a wide range of dust observations in the local universe. In particular, the relation between galactic D/Z and metallicity along with its scatter, gas-phase element depletion trends, and varying dust population chemical compositions seen in the MW and Large and Small Magellanic clouds \citepalias{choban_2024:DustyLocaleEvolution}.

\begin{table}
	\centering
        \renewcommand{\arraystretch}{1.3}
        \begin{tabular}{| l | c  c |}
            \hline
            Physical Quantity & Fiducual & Enhanced \\ [0.5ex] 
            \hline\hline
                $\tau_{\rm  g,sil}^{\rm CNM/MC}$ (Myr) & 27.5/8.2 & 6.9/2.0 \\
                $\tau_{\rm  g,carb}^{\rm CNM/MC}$ (Myr) & 113/$\infty$ & 28.3/$\infty$ \\
                $\tau_{\rm  g,iron}^{\rm CNM/MC}$ (Myr) & 1.7/7.7 & 0.4/1.9 \\
                $M_{\rm  SNe\,II,sil}$ ($\Msol$) & $1.7 \times 10^{-4}$ & 0.095 \\
                $M_{\rm  SNe\,II,carb}$ ($\Msol$) & 0.020 & 0.027 \\
                $M_{\rm  SNe\,II,iron}$ ($\Msol$) & $7.4\times 10^{-5}$ & 0.015 \\
                \hline		
        \end{tabular}
	\caption{Quantitative comparison of the gas-dust accretion timescales and SNe II dust yields for our fiducial and enhanced models for silicates, carbonaceous, and metallic iron dust species. The typical accretion timescales are provided for CNM gas (e.g. $\nH=30$ cm$^{-3}$, $T=100$ K, $Z=0.1\Zsol$) and MC gas (e.g. $\nH=10^3$ cm$^{-3}$, $T=30$ K, $Z=0.1\Zsol$) assuming \citet{asplund_2009:ChemicalCompositionSun} element abundances. Note carbonaceous dust does not grow via accretion in MC environments due to the lock up of C into CO (see Appendix B in \citealt{choban_2022:GalacticDustmodellingDust}). Our fiducial model SNe II yields are taken from \citet{zhukovska_2008:EvolutionInterstellarDust}. We highlight that the theoretical maximum total SNe II dust yield (i.e. assuming all Si, C, and Fe are locked in dust and no dust destroyed by the SNe reverse shock) given the \citet{nomoto_2006:NucleosynthesisYieldsCorecollapse} metal yields used in FIRE-2 is $M_{\rm SNe\,II, max}=0.685 \Msol$.}
	\label{tab:model_comparison}
\end{table}

We also reran all of our simulations with an `enhanced' version of our dust evolution model to investigate how uncertainties in the dust life cycle could affect our results. In particular, the `enhanced' model increases the SNe II dust creation efficiency to $20\%$\footnote{Our model's definition of SNe creation efficiency is the fraction of key element locked into dust. So a silicate creation efficiency specifies the fraction of Si locked into silicate, with the corresponding amounts of O, Mg, Fe determined by the assumed silicate chemical composition as specified in \citetalias{choban_2022:GalacticDustmodellingDust}.} for all dust species and decreases the gas-dust accretion timescale by a factor of 4.
We label these two model versions as Fiducial and Enhanced in Table~\ref{tab:simulations} and use the {\bf \_enh} suffix to distinguish individual simulations run with the `enhanced' version.
The reasoning for the changes used in the `enhanced' model are discussed in detail in Sec~\ref{sec:dust_model}, and we provide a brief summary below. {\bf (1)} There are large theoretical and observational uncertainties in SNe II dust production, with SNe II creation efficiencies of $20\%$ being on the upper end of observations \citep[e.g.][]{shahbandeh_2023:JWSTObservationsDust}. {\bf (2)} Our dust-gas accretion routine does not include all physical processes, such as dust-gas clumping factor and grain size evolution, and thus could overpredict accretion timescales.
We also showcase other tested changes to the dust model for one galaxy in Appendix~\ref{app:dust_evo_variations}.
For ease of comparison we provide the typical gas-dust accretion timescales for cold neutral medium (CNM; e.g. $\nH=30$ cm$^{-3}$, $T=100$ K, $Z=0.1\,\Zsol$) and molecular cloud (MC; e.g. $\nH=10^3$ cm$^{-3}$, $T=30$ K, $Z=0.1\,\Zsol$) gas and SNe II dust yields for silicates, carbonaceous, and metallic iron dust used in the fiducial and `enhanced' models in Table~\ref{tab:model_comparison}. We stress that the accretion timescales provided are not fixed values, instead depending on the local gas properties.

We highlight that there are some differences between the final redshifts and resulting galactic properties presented here and those from the original simulation runs listed in \citet{ma_2019:DustAttenuationDust}. These differences are due to changes in FIRE-2 and how final redshifts were originally chosen. Specifically, the simulations in \citet{ma_2019:DustAttenuationDust} were run with a version of FIRE-2 that included erroneous heating from the cosmic ray background at high redshift. This extra heating suppressed star formation in the intergalactic medium, restricting it to the densest regions in the galactic halo \citep{su_2018:DiscreteEffectsStellar,garrison-kimmel_2019:StarFormationHistories}. We find that the removal of this extra heating has little effect on the star formation histories of each galaxy, but star formation is generally more dispersed, occurring in less dense regions as well, but still above the threshold density for star formation. The main consequence of this dispersed star formation is a significant slowdown of the simulation, primarily for the earliest forming massive galaxies. This then affects the `final' redshift of the simulations, which were originally chosen to be the point at which the simulations became too computationally expensive to continue.
The end result is a higher `final' redshift for simulations originally run to $\zfinal=7$ and $9$, and slight variations between reruns of the same simulation.

\begin{figure*}
    \plotsidesize{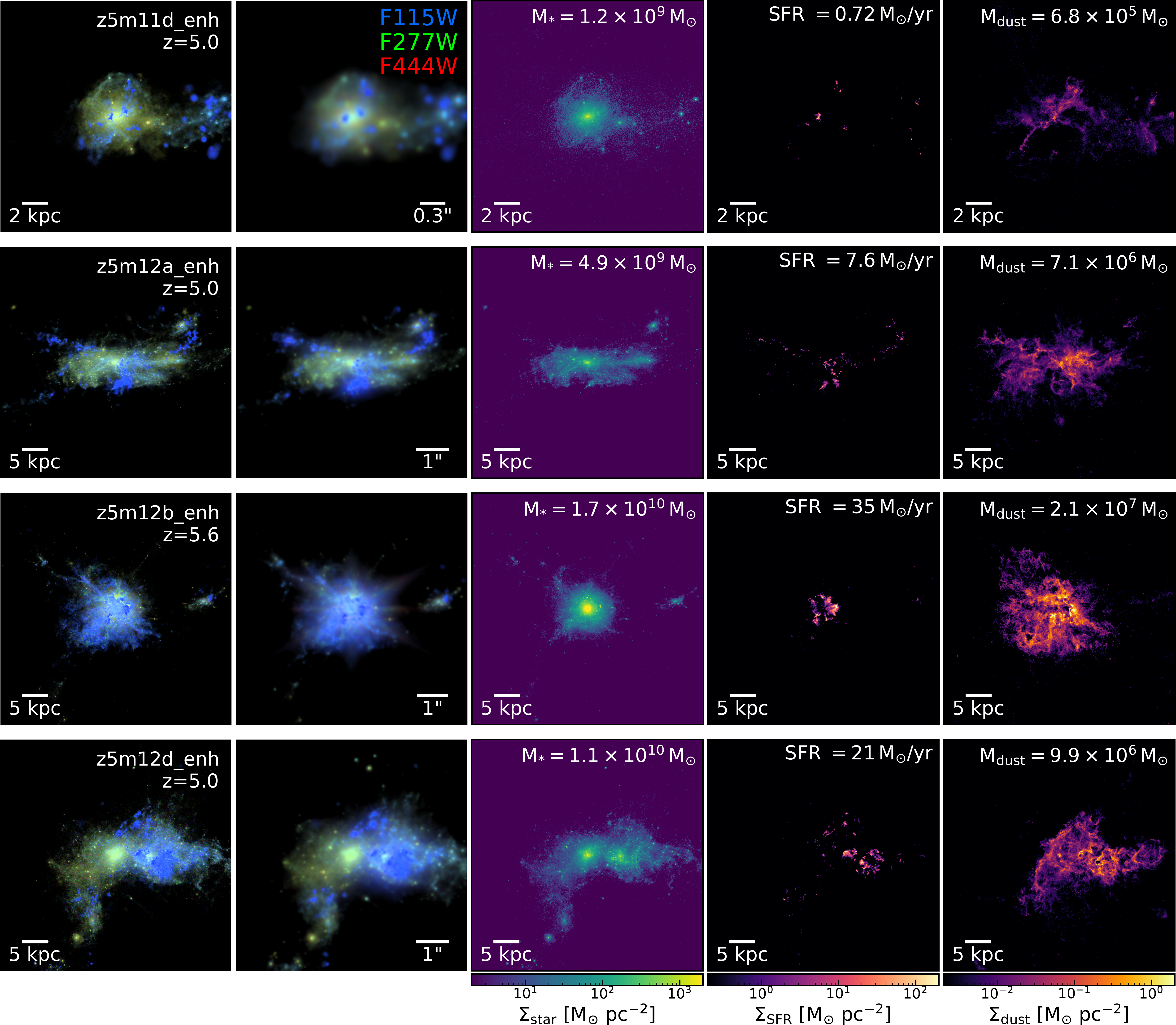}{0.95}
    \vspace{-0.25cm}
    \caption{Images and projections of galaxies {\bf z5m11d}, {\bf z5m12a}, {\bf z5m12b}, and {\bf z5m12d} at $\zfinal$ (see Table~\ref{tab:simulations}) run with our `enhanced' dust evolution model (note the differences in scale). ({\it left}) High-resolution, noise-free observer-frame JWST RGB image using NIRCam F115W, F277W, and F444W filters, showing three orders of magnitude in surface brightness and following the \citet{lupton_2004:PreparingRedGreenBlueImages} color algorithm. This image was made with SKIRT utilizing the local D/Z produced by our model along with assumed SMC dust opacities. ({\it middle left}) The same image down-sampled to the resolution of NIRCam (0".031 pixel$^{-1}$) and convolved with each filter's PSF. ({\it middle}) Stellar surface density projection. ({\it middle right}) SFR surface density projection for stars ${<}10$ Myr old. ({\it right}) Dust surface density projection. These images highlight the complex and varying morphologies of galaxies in our suite. The $\zfinal\sim5$ galaxies have varying shapes, ranging from compact to extended, due to their bursty star formation. All galaxies exhibit prominent dusty structures in the inner halo.}
    \label{fig:mock_JWST1}
\end{figure*}

\begin{figure*}
    \plotsidesize{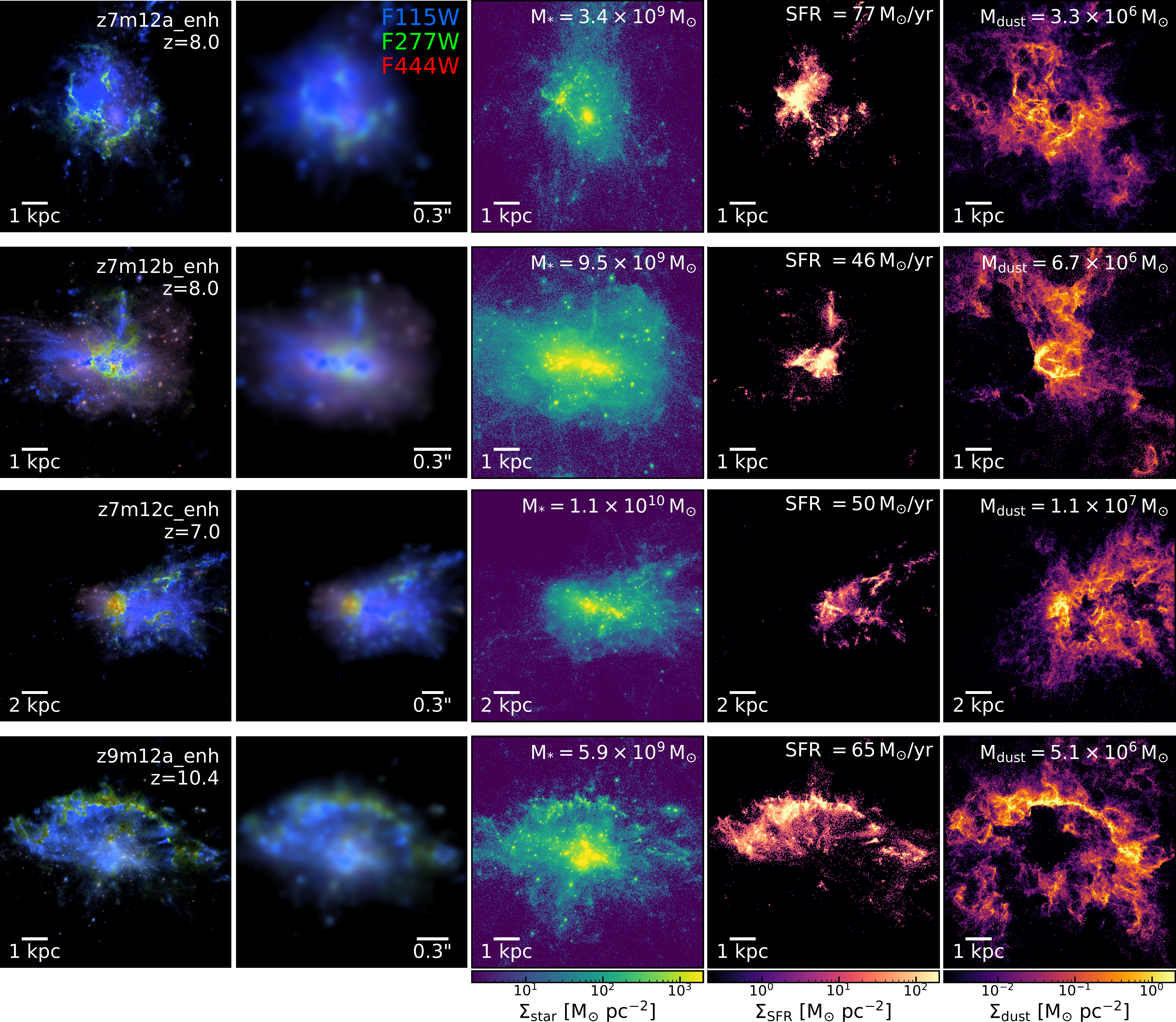}{0.95}
    \vspace{-0.25cm}
    \caption{Same as Fig.~\ref{fig:mock_JWST1} for galaxies {\bf z7m12a}, {\bf z7m12b}, {\bf z7m12c}, and {\bf z9m12a} at $\zfinal$ run with the `enhanced' dust evolution model.}
    \label{fig:mock_JWST2}
\end{figure*}

\subsection{Model Limitations} \label{sec:model_limitations}

Theoretical models cannot incorporate every important physical process due to resolution and computational constraints, and our models are no different. Below, we list the main physics limitations of our galaxy and dust evolution models in regards to missing/excluded physics.

{\bf AGN:} Our galaxy formation model does not include the formation and feedback of active galactic nuclei (AGN) nor does our dust evolution model include the possible nucleation of dust grains in AGN outflows \citep{elvis_2002:SmokingQuasarsNew}.
However, AGN are unlikely to form in the types of galaxies we simulate \citep{harikane_2023:JWSTNIRSpecFirst,maiolino_2023:JADESDiversePopulation,matthee_2024:LittleRedDots} and the current understanding of AGN dust nucleation is highly uncertain \citep[][see Sec. 6]{schneider_2024:FormationCosmicEvolution}.

{\bf Cold ISM:} While our simulations resolve the multi-phase ISM, an advantage compared to previous theoretical works discussed in Sec.~\ref{sec:simulation_comparison}, the predicted structure of the cold, dense ISM,  especially the dense molecular phase \citep{keating_2020:ReproducingCOH$_2$Conversion}, is physics and resolution-dependent. This then affects our dust evolution model's predictions for the gas-dust accretion process, which occurs in the cold neutral and molecular media. To test our model's sensitivity to variations in ISM evolution and structure, we compare the predicted results of the `enhanced' dust evolution model for one galaxy run with the FIRE-2 and FIRE-3 \citep{hopkins_2023:FIRE3UpdatedStellar} stellar feedback and ISM physics models in Appendix~\ref{app:dust_evo_variations}. 
Notably, FIRE-3 includes improved modeling of cold gas from the CNM to molecular clouds and removes the density threshold criteria for star formation. 
Overall we find the predicted $\Mdust$ and median D/Z are slightly lower with FIRE-3 due to a ${\lesssim}50\%$ reduction in cold ($T\leq1000$ K) gas mass.

{\bf Dust Physics:} We note that all cooling and heating processes and radiative transfer modeled in our simulations are not coupled with our dust evolution model and instead follow the default assumptions in FIRE-2. Specifically, dust heating and cooling and radiative transfer assume a constant D/Z ratio, and metal-line cooling assumes no metals are locked in dust. 
This choice was made to avoid possible changes to galaxy evolution and resulting $\zfinal$ galactic properties, which could affect our dust evolution predictions.
In future work, we will investigate what effects the full integration of our dust evolution model with FIRE physics has on predicted galaxy evolution.

{\bf Grain Size:} The primary limitation of our dust evolution model is the exclusion of evolving grain sizes.
Since most dust processes depend on a dust population's effective grain size, our model's predictions will be weakest in environments where the dust population is expected to deviate significantly from an MRN \citep{mathis_1977:SizeDistributionInterstellar} grain size distribution.
Noteably, a reduction in small grains will reduce the efficiency of gas-dust accretion, dust destruction by SNe, and thermal sputtering. Grains created by SNe are observed to be much larger than expected from an MRN distribution \citep{gall_2014:RapidFormationLarge,wesson_2015:TimingLocationDust,bevan_2016:ModellingSupernovaLine,priestley_2020:DustMassesGrain} and in MC environments coagulation efficiently converts small grains into large grains \citep{hirashita_2009:ShatteringCoagulationDust}. However, large grains shatter into smaller grains on short (${<}5$ Myr) timescales in the WIM \citep{hirashita_2009:ShatteringCoagulationDust}. Thus, gas-dust accretion and thermal sputtering are likely unaffected since they are most efficient in the CNM and hot ionized medium respectively. Grains are also efficiently shattered in SNe shocks \citep{kirchschlager_2019:DustSurvivalRates}, and so this is unlikely to affect dust destruction by SNe beyond the many other uncertainties that already exist in this processes \citep{kirchschlager_2023:DustSurvivalRates,kirchschlager_2024:TotalDestructionComplete}.

\section{Results} \label{sec:results}

We first showcase mock observer-frame JWST composite images along with stellar and dust surface density projections for each galaxy to highlight the breadth of galaxy morphologies and variable dust structures contained in our simulation suite.
Fig.~\ref{fig:mock_JWST1} and~\ref{fig:mock_JWST2} show JWST images of all galaxies at $\zfinal$ with our `enhanced' dust evolution model created using the radiative transfer code SKIRT \citep{camps_2015:SKIRTAdvancedDust}. 
These images use STARBURST99 \citep{leitherer_1999:Starburst99SynthesisModels} to compute the stellar spectra for each star particle given their age and metallicity and use the tracked dust mass for each gas cell produced by our dust model assuming a SMC dust population \citep{weingartner_2001:DustGrainSizeDistributions}.
We showcase noise-free high-resolution RGB composite images using JWST NIRCam F115W, F277W, and F444W filters (similar to the JADES survey; \citealt{rieke_2023:JADESInitialData}) convolved over each ﬁlter's transmission curve.
We also show the same RGB composite images down-sampled to the resolution of NIRCam (0".031 pixel$^{-1}$) and convolved with each filters' PSF provided by {\small webbpsf} \citep{perrin_2012:SimulatingPointSpread}.
Projections of each galaxy's stellar mass, star formation, and dust mass surface density are 
also provided for reference.

The $\zfinal\sim5$ galaxies vary in morphology, from compact systems with branches of infalling gas ({\bf z5m11d} and {\bf z5m12b}) to extended complexes with prominent substructures ({\bf z5m12a} and {\bf z5m12d}). These variations largely arise from bursts in star formation that evacuate large amounts of gas from the galactic center, shutting down star formation until gas recollapses back onto the center, triggering the next burst. The $\zfinal\sim7$ and 9 galaxies are uniform in their heterogeneity, having complex and extended structures due to the chaotic interactions of multiple infalling systems that have yet to fully settle into one central galaxy. Regardless of these differences, all galaxies exhibit dusty structures at or around their centers.

\begin{figure*}
    \centering
    \plotsidesize{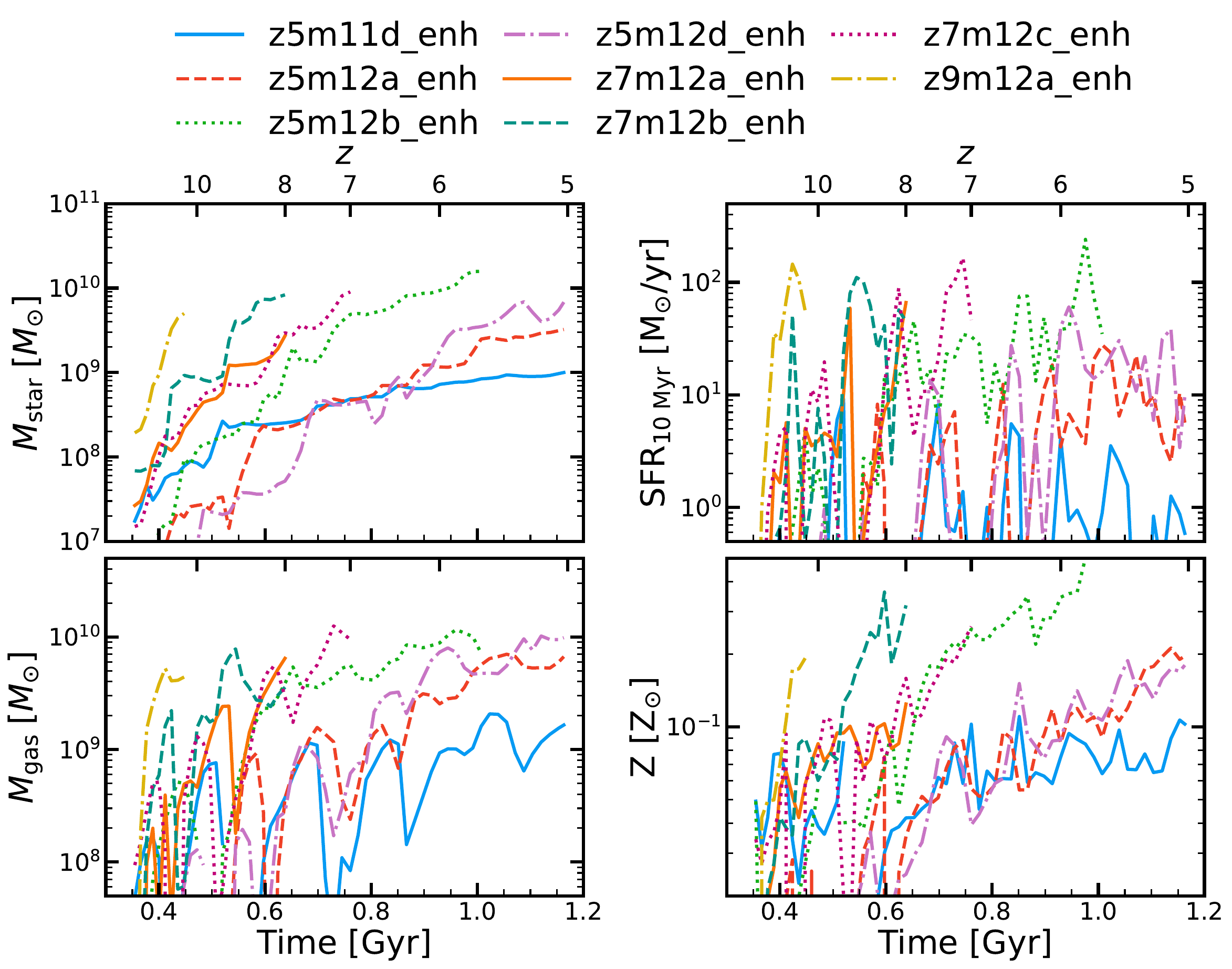}{0.85}
    \vspace{-0.25cm}
    \caption{Evolution of total stellar mass ({\it top left}), star formation averaged over 10 Myr intervals ({\it top right}), gas mass ({\it bottom left}), and median metallicity in cool ($T<1000$ K) gas ({\it bottom right}) within 0.2$\Rvir$ for our simulated galaxies. For clarity, we only include one simulation for each galaxy run with the `enhanced' dust evolution model since reruns of the same galaxy have near identical evolution.
    We also note that the short drops in stellar mass are due to major merger events during which the galactic center jumps between the two merging galaxies.
    The resulting metallicities for our galaxies are quite low compared to present-day stellar-mass analogs. These low metallicities delay the onset of efficient accretion, which is determined by a critical metallicity threshold, to later points in the galaxy's life and reduce the rate of dust growth via accretion, both critical for the buildup of dust mass. Furthermore, as is evident by the large drops in $M_{\rm gas}$ following peaks in SFR, these galaxies are extremely burst, experiencing multiple rapid rises in their star formation, which evacuate large amounts of gas, metals, and dust from the inner part of the halo. This disruption of cold gas on a galactic scale delays the build-up of dust mass in the galaxy since accretion only occurs in such environments.}
    \label{fig:enhanced_galaxy_evolution}
\end{figure*}

\subsection{Evolution of Galactic Properties} \label{sec:galaxy_evolution}

In Fig.~\ref{fig:enhanced_galaxy_evolution}, we present the evolution of various galactic properties, specifically the stellar mass, star formation rate averaged over 10 Myr intervals, gas mass, and mass-weighted median metallicity of cool ($T<1000$ K) gas for each galaxy. These values are determined from star particles and gas cells within $0.2\Rvir$ of the galactic center for all galaxies\footnote{We choose $0.2\Rvir$ as the outer boundary for our galaxies due to the tendency of high-redshift galaxies to have more expansive stellar populations relative to their virial radii as compared to galaxies at lower redshift, where $0.1\Rvir$ is commonly used. This
is also consistent with other works that analyze the same simulation suite \citep{sun_2023:SeenUnseenBursty,sun_2023:BurstyStarFormation,marszewski_2024:HighRedshiftGasPhaseMass} and we find only a small amount of the stellar mass within the virial halo is excluded by this cutoff, except for during mergers.}.
For clarity, we only show simulations run with the `enhanced' dust evolution model.
While there are stochastic variations between simulations of the same galaxy, the overall evolution is very similar as can be seen in Appendix~\ref{app:dust_evo_variations}. 
We highlight two galactic properties that are critical for the buildup of galactic dust mass. 

{\bf (1)} Metallicity: The galactic metallicity determines both the rate of dust buildup via accretion of gas-phase metals onto preexisting dust grains (accretion timescale inversely scales with the local metallicity) and when dust growth via accretion becomes more efficient than dust destruction via SNe (above a given $\Zcrit$). Therefore, galaxies that reach higher metallicities earlier will experience a faster build-up of their dust mass and have more time to do so.
As seen in Fig.~\ref{fig:enhanced_galaxy_evolution}, the metallicity of our galaxies is low compared to present-day stellar mass analogs\footnote{The normalization of the MZR in FIRE-2 decreases by ${\sim}0.4$ dex from
$z = 0 - 3$ and evolves weakly for $z\gtrsim3$ \citep{bassini_2024:InflowOutflowProperties,marszewski_2024:HighRedshiftGasPhaseMass}.}, with cool, dense gas reaching $Z\sim0.4 \Zsol$ for the most metal-rich galaxy in our sample. 
Some galaxies, such as {\bf z7m12b} and {\bf z9m12a}, reach these high metallicities earlier than others due to high SFR bursts (${\sim}10^2 \Msol/$yr) at early times, suggesting they are more likely to be dust-rich.

{\bf (2)} Burstiness: Bursts in star formation can lead to galaxy-wide blow outs of cold gas where gas-dust accretion occurs ($T < 300$ K for our model), hindering dust growth and ultimately slowing the buildup of dust.
As evident from their star formation histories in Fig.~\ref{fig:enhanced_galaxy_evolution}, all galaxies in our suite are extremely bursty, experiencing multiple, rapid increases in their SFR. The resulting feedback from these bursts evacuates large amounts of gas, metals \citep{muratov_2015:GustyGaseousFlows,muratov_2017:MetalFlowsCircumgalactic}, and dust from the inner halo sometimes removing all gas from $0.2 \Rvir$.

\begin{figure*}
    \centering
    \plotsidesize{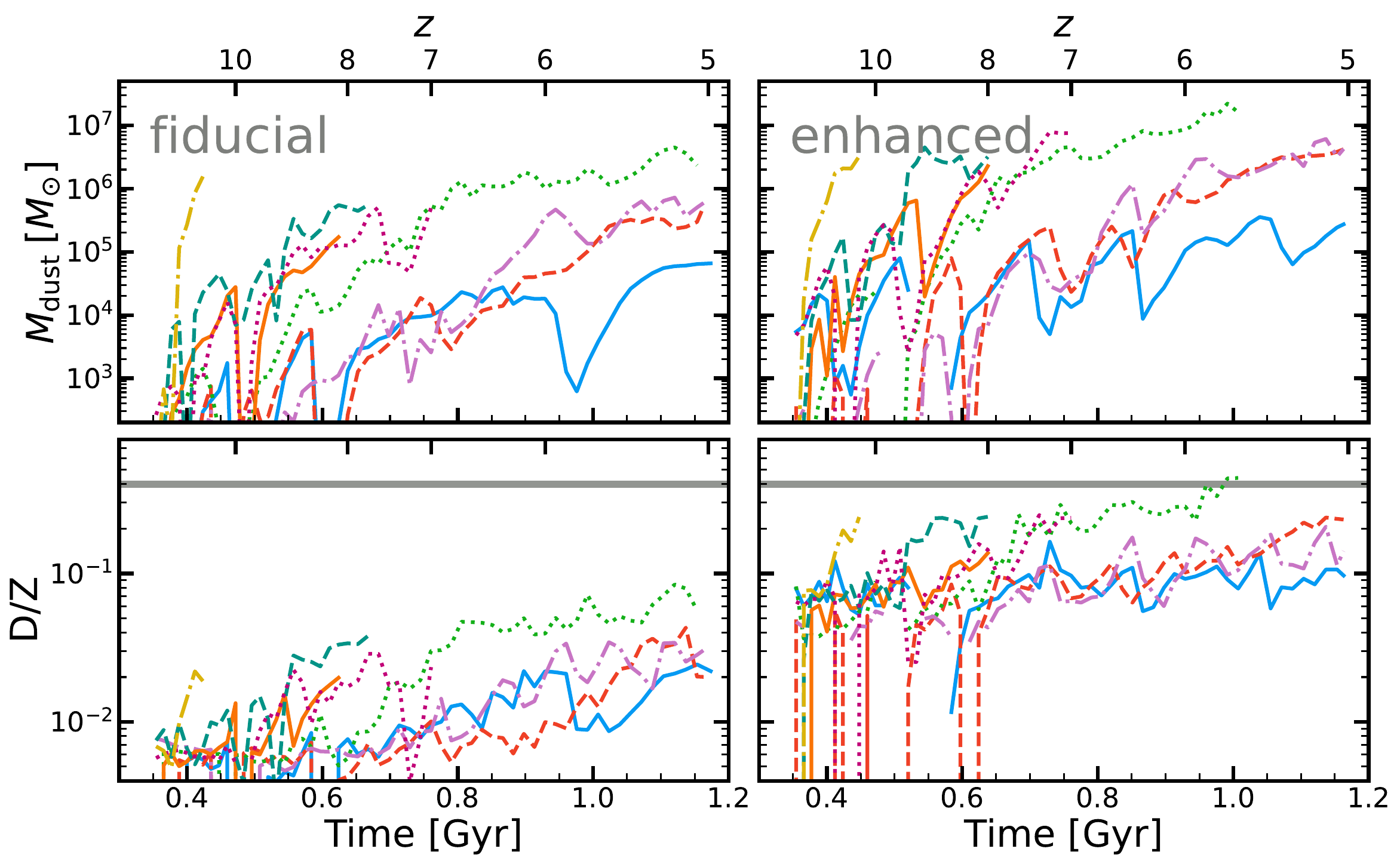}{0.9}
    \vspace{-0.25cm}
    \caption{Evolution of the total dust mass ({\it top}) and median D/Z in cold neutral ($T<1000$ K) gas ({\it bottom}) within $0.2\Rvir$ for our simulated galaxies produced by the fiducial ({\it left}) and `enhanced' ({\it right}) dust evolution models. Galaxy labels are the same as Fig.~\ref{fig:enhanced_galaxy_evolution}. We also highlight the `canonical' \citet{draine_2007:InfraredEmissionInterstellar} MW D/Z $=0.4$ ({\it grey line}). While both models initially predict low dust masses determined by the SNe II dust creation efficiency, the fiducial model produces ${\sim}1$ dex lower dust masses. Eventually, the galactic dust masses begin to increase due to the onset of efficient dust growth via accretion. However, for the fiducial model, this buildup occurs later and is slower on average due to longer accretion timescales resulting in higher critical metallicity thresholds. Ultimately, the fiducial model produces ${\lesssim}1$ dex lower dust masses when compared to the `enhanced' model.}
    \label{fig:Mdust_evolution}
\end{figure*}

\begin{figure*}
    \centering
    \plotsidesize{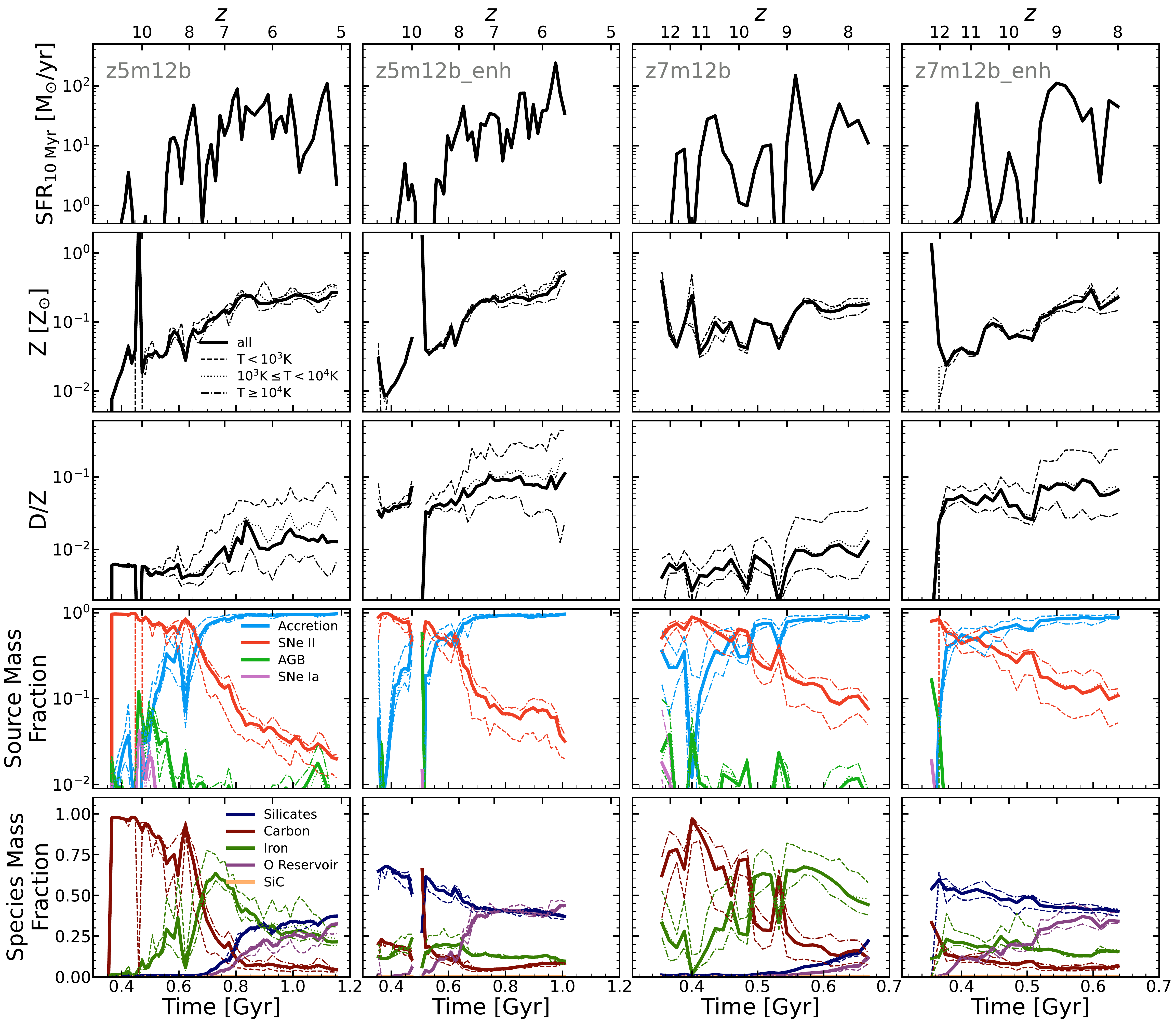}{0.99}
    \vspace{-0.25cm}
    \caption{Time evolution of SFR averaged over last 10 Myr (\textit{top}), median metallicity (\textit{second from top}), D/Z (\textit{middle}), dust creation source mass fraction (\textit{second from bottom}), and dust species mass fraction (\textit{bottom}) for {\bf z5m12b} and {\bf z7m12b} run with the fiducial (\textit{left/middle right}) and `enhanced' (\textit{middle left/right}) dust evolution models. Median values are given for all gas within $0.2\Rvir$ of the galactic halo (\textit{solid}), cold neutral gas (\textit{dashed}), warm neutral gas (\textit{dotted}), and warm/hot ionized gas (\textit{dash-dotted}). Note the different x-axis scales for {\bf z5m12b} and {\bf z7m12b}. We also highlight the spikes in Z are due to blowout events where all or almost all gas mass is ejected outside of $0.2\Rvir$, as can be seen in Fig.~\ref{fig:enhanced_galaxy_evolution}, with the majority of gas left being hot, metal-rich SNe ejecta.
    For both models, the galactic dust populations are initially dominated by SNe II created dust but differ in dust amount and chemical composition. In particular, the fiducial model predicts almost entirely carbonaceous dust with D/Z $<0.01$, while the `enhanced' model predicts predominantly silicate dust, with some carbonaceous and metallic iron, and D/Z $\sim0.05$. Eventually, the galactic metallicity reaches $\Zcrit$ for first metallic iron and later silicates, causing accretion to be the dominant producer of dust mass and increasing the median D/Z. The rise in D/Z once accretion becomes dominant is slightly faster for the `enhanced' model, and reaches a higher median D/Z in cold, neutral gas, D/Z $\sim0.3$ compared to D/Z $\sim0.06$ for the fiducial model. Both models also predict a large 1 dex spread in D/Z across gas phases despite only a factor of ${\lesssim}2$ difference in metallicity. }
    \label{fig:enhanced_dust_evolution}
\end{figure*}

The compounding effects of low metallicity and burstiness can be seen in the evolution of the galactic dust mass and dust population composition. 
In Fig.~\ref{fig:Mdust_evolution}, we compare the resulting evolution of the galactic dust mass and median D/Z produced by the fiducial and `enhanced' dust evolution model for all galaxies in our suite. 
In Fig.~\ref{fig:enhanced_dust_evolution}, we highlight the dustiest galaxies in the $\zfinal\sim5$ and $7$ subsets, {\bf z5m12b} and {\bf z7m12b}, comparing a detailed breakdown of each galaxy's metal and dust population evolution produced by the two dust evolution models.
We include the median metallicity, median D/Z, dust creation source contribution, and dust species composition. We also include the breakdown for each gas phase: cold neutral gas ($T<10^3$ K), warm neutral gas ($10^3$ K $\leq T < 10^4$ K), and warm/hot ionized gas ($T\geq10^4$ K). Again, only gas within $0.2\Rvir$ of the galactic center is considered.
We provide a similar comparison for all other galaxies in our simulation suite in Appendix~\ref{app:addtional_figs}.

For the fiducial model, galaxies are initially extremely dust-poor (D/Z $<0.01$; $\Mdust\sim10^3\Msol$) and entirely dominated by carbonaceous dust due to an assumed low SNe II dust creation efficiencies for all species besides carbonaceous as shown in Table~\ref{tab:model_comparison}. Accretion eventually takes over as the dominant dust producer as a galaxy's median metallicity reaches $\Zcrit$ for first metallic iron and possibly later silicates, growing each successively as seen by the increase in their respective species mass fractions.
This buildup is relatively quick (${\sim}0.2$ Gyr), but the equilibrium values achieved are quite low with a maximum $\Mdust\sim3\times10^6\Msol$ and median D/Z $\sim0.08$ in cold neutral gas for our dustiest galaxy at $\zfinal$. Overall, the fiducial model struggles to build up a sizable dust mass for all galaxies in our simulation suite due to low SNe II dust creation efficiencies coupled with possibly overestimated accretion timescales, as we discuss in Sec.~\ref{sec:dust_model}.
In contrast, the `enhanced' model produces larger initial dust masses (D/Z $\sim0.05$; $\Mdust\sim5\times10^4\Msol$) dominated by silicate dust, with appreciable amounts of carbonaceous and metallic iron dust, due to the increased SNe II dust creation efficiencies for all dust species.
Similar to the fiducial model, accretion eventually takes over as the dominant producer of dust mass. However, the onset of efficient dust growth occurs at earlier times (i.e. lower $\Zcrit$), and the build-up is faster due to the model's $4\times$ decrease in accretion timescales for all dust species.
Metallic iron is the first dust species to grow, and then silicates, and finally carbonaceous (in the case of {\bf z5m12b}) as can be seen in the increase of the species mass fraction or metallic iron, O Reservoir\footnote{The O Reservoir dust species is essentially a bucket that sequesters excess O to match observed O depletions in the MW, with the amount of extra O sequestration scaling with the fraction of the maximum formable amount of silicate dust present in a gas cell (see Sec. 2.3.2 in \citetalias{choban_2022:GalacticDustmodellingDust} for details). Thus, when silicate dust grows, the O Reservoir species also grows.}, and carbonaceous, respectively. Ultimately this leads to ${\lesssim}1$ dex higher dust mass for all galaxies with a maximum $\Mdust \sim 10^7 \Msol$ and median D/Z $\sim 0.3$ in the cool neutral gas of our dustiest galaxy at $\zfinal$. 
Another notable result shared by both models is a large spread in D/Z, up to 1 dex, across gas phases. This is particularly relevant for works that post-process simulations with an assumed dust population which we discuss in Sec.~\ref{sec:simulation_comparison}.

\subsection{Comparisons and Predictions for $z \gtrsim 5$ Observations} \label{sec:results_observations}

\begin{figure*}
    \centering
    \plotsidesize{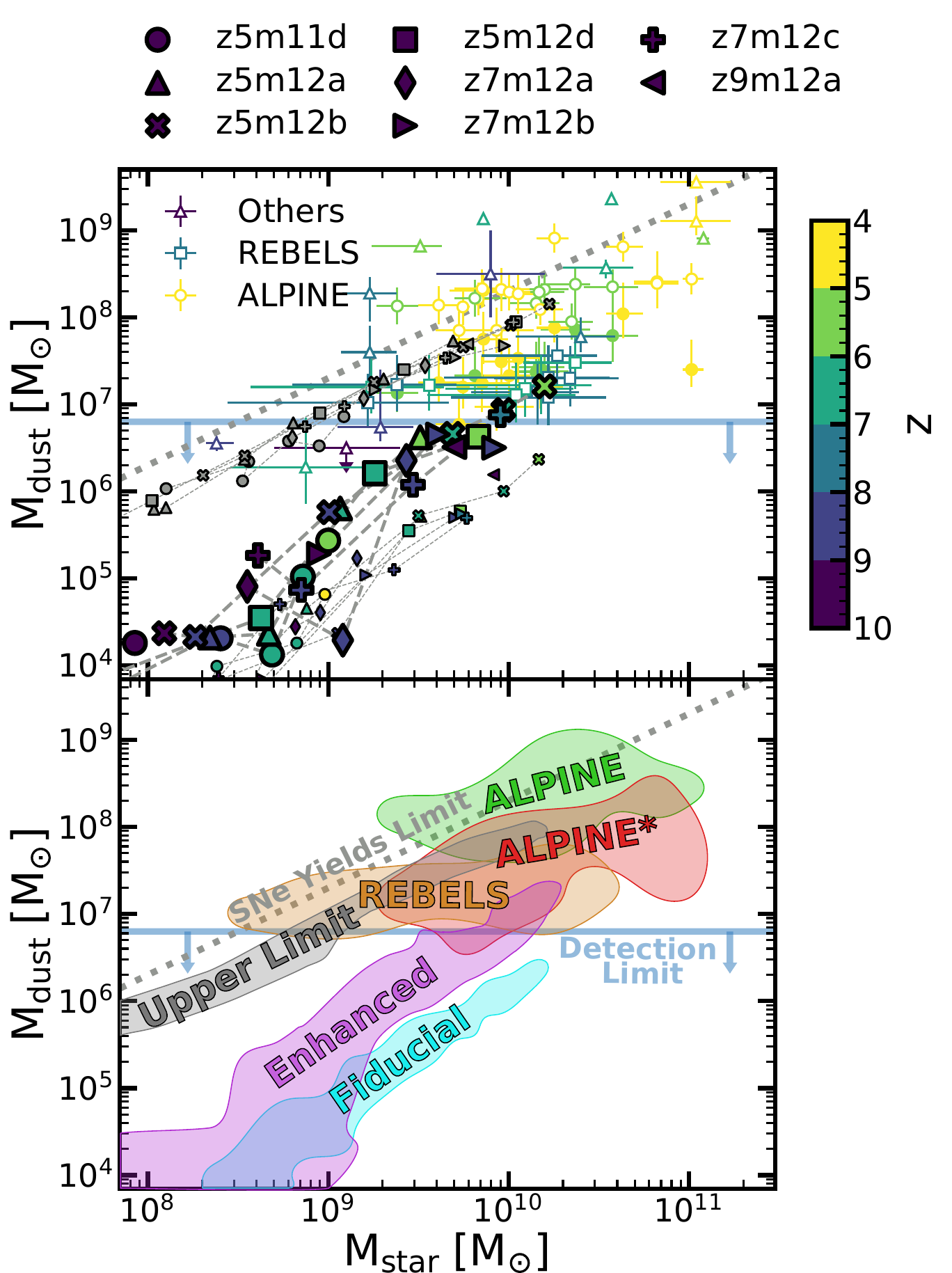}{0.6}
    \vspace{-0.25cm}
    \caption{Resulting relation between galactic dust mass and stellar mass within $0.2\Rvir$ of the galactic center for our galaxies. We include ({\it top}) the exact results from our simulations and observations and ({\it bottom}) a schematic representation for ease of reading which shows shaded regions encompassing our simulation results and observational samples with error bars. We present results from our simulations run with the fiducial ({\it small colored}) and `enhanced' ({\it large colored}) dust evolution model. Each connected point is -1 redshift apart starting at $z=10$, including an additional point at the simulation's final redshift, with the color denoting the redshift. We also provide two upper limits on the possible dust mass. 
    One ({\it small grey}) is an upper limit for each galaxy assuming a constant D/Z $=0.6$ for all gas within $\Rvir$. Note the upper bound stellar masses increase slightly due to the change in radius. 
    The other ({\it grey dashed}) is an upper limit determined by the total amount of metals produced by a galaxy of a given stellar mass assuming the same IMF and SNe yields as used in FIRE-2 ($M_{\rm metals}\sim0.02\Mstar$). 
    We include observations from ALMA surveys ALPINE, from \citet{faisst_2020:ALPINEALMAIISurvey,schaerer_2020:ALPINEALMAIISurvey,pozzi_2021:ALPINEALMACIISurvey}, ({\it hollow circles}) and REBELS, from \citet{inami_2022:ALMAREBELSSurvey,topping_2022:ALMAREBELSSurvey,sommovigo_2022:ALMAREBELSSurvey}, ({\it hollow squares}) and group together numerous small number observations from \citet{watson_2015:DustyNormalGalaxy}, \citet{laporte_2017:DustReionizationEra}, \citet{strandet_2017:ISMPropertiesMassive}, \citet{tamura_2019:DetectionFarinfraredIii}, \citet{witstok_2023:CarbonaceousDustGrains}, \citet{akins_2023:TwoMassiveCompact}, \citet{valentino_2024:ColdInterstellarMedium}, and one $z=10.6$ upper limit estimate from \citet{fudamoto_2024:NOEMAObservationsGNz11} ({\it hollow triangles}). Alternate ALPINE $\Mdust$ estimates from \citet{sommovigo_2022:NewLookInfrared} are also included ({\it filled circles}; ALPINE* in schematic). An approximate $\Mdust$ detection threshold for the REBELS program ({\it blue line}) similar to \citep{sommovigo_2022:NewLookInfrared}. Note the ALPINE program detection threshold is only ${\sim}0.1$ dex higher.
    Both models predict a roughly constant relation with redshift with ${\sim}0.5$ dex scatter at any given stellar mass. However, the fiducial model is unable to reproduce the large dust masses observed at any stellar mass or redshift. The `enhanced' model is able to reproduce observed dust masses for the $\Mstar\sim10^{10}\Msol$ end of the REBELS $z\sim7$ sample, but falls ${\sim}1$ dex below the ALPINE $z\sim5$ sample. However, the $\Mdust$ from the ALPINE sample may be overpredicted, as discussed in Sec.~\ref{sec:observations}. Moreover, given the metal content of our galaxies, only an unrealistic assumption of D/Z $=0.6$ everywhere in the galactic halo could produce $\Mdust$ value similar to those in the ALPINE sample.}
    \label{fig:Mdust_Mstar_relation}
\end{figure*}

\begin{figure}
    \centering
    \includegraphics[width=0.99\columnwidth]{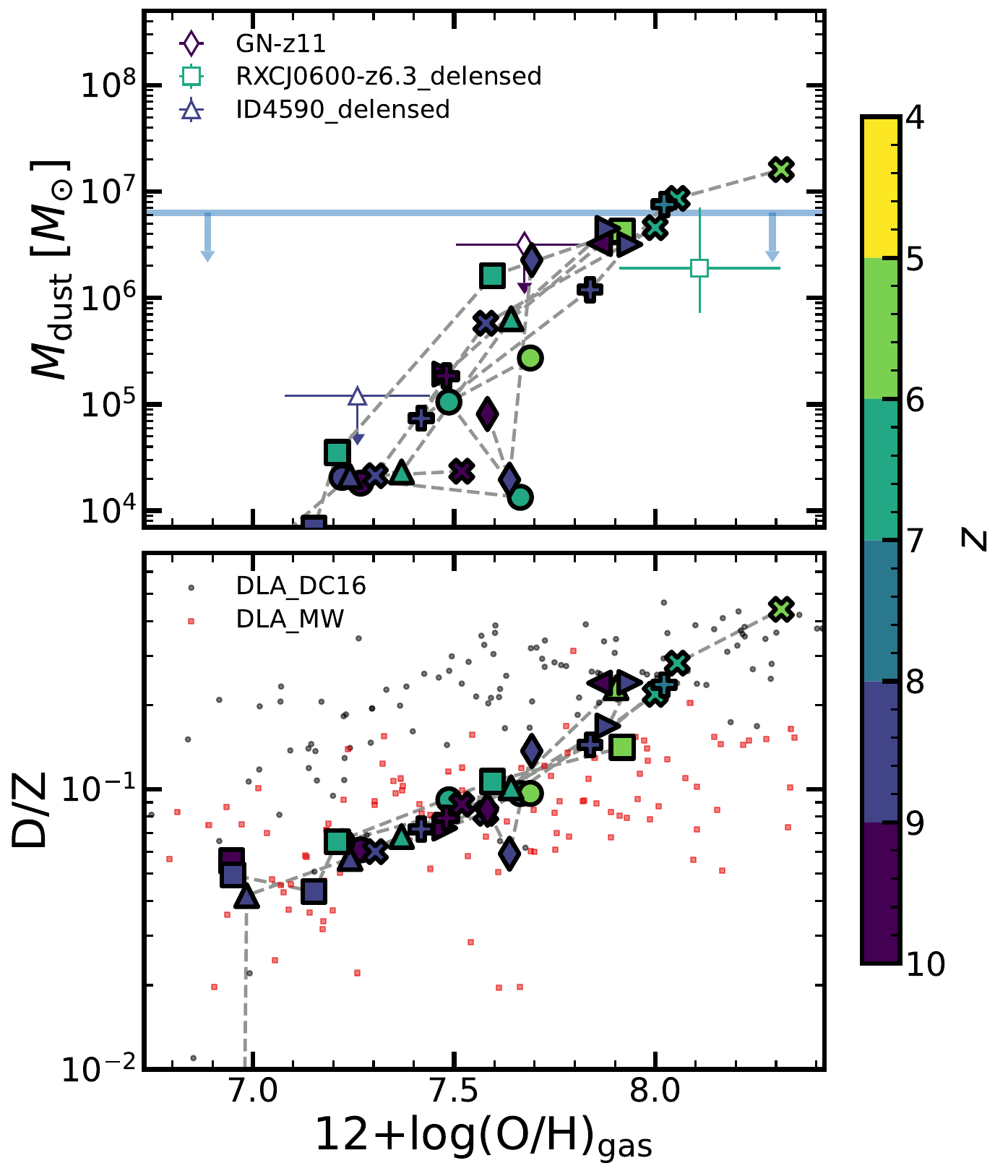}
    \vspace{-0.25cm}
    \caption{Resulting relation between galactic dust mass ({\it top}) and median D/Z for cool, neutral ($T<1000$ K) gas ({\it bottom}) with respect to median oxygen abundance for simulations run with our `enhanced' model. Our oxygen abundance definition only considers gas-phase oxygen (only O not depleted into dust) for gas with $7000<T<15000$ and $\nH>0.5$ cm$^{-3}$, similar to regions probed by nebular emission lines. We include observations of galaxies with both metallicity and dust mass estimates or upper limits. Notably, GN-z11 \citep{bunker_2023:JADESNIRSpecSpectroscopy,fudamoto_2024:NOEMAObservationsGNz11} and lensed galaxies RXCJ0600-z6.3 \citep{fujimoto_2024:JWSTALMAMultipleline,valentino_2024:ColdInterstellarMedium} and ID4590 \citep{heintz_2023:GasStellarContent,fujimoto_2024:JWSTALMAMultipleline}. We also include observed D/Z versus metallicity measurements for DLAs out to $z\sim5$ from \citet{decia_2016:DustdepletionSequencesDamped, quiret_2016:ESOUVESAdvanced} using [Zn/Fe] calibrations from \citet{decia_2016:DustdepletionSequencesDamped} ({\it black circles}) and MW-derived calibrations from \citet{roman-duval_2022:METALMetalEvolution} ({\it red squares}). Note SMC and LMC-derived calibration fall between the two.
    Galaxy labels, redshift colors, and $\Mdust$ detection threshold are the same as Fig.~\ref{fig:Mdust_Mstar_relation}.}
    \label{fig:dust_Z_relations}
\end{figure}

Observations of galactic dust properties at $z\gtrsim 5$ are limited in quantity, with only ${\sim}100$ direct detections of galactic rest-frame IR dust continuum. 
Furthermore, a majority of these observations provide relatively limited dust population information, usually only $\Mdust$, although JWST is beginning to deliver dust attenuation estimates \citep[e.g.][]{witstok_2023:CarbonaceousDustGrains}.
Due to this limited data, the most commonly examined relation at these high redshifts is the relation between the galactic dust mass and stellar mass (dust-to-stellar mass ratio; $\Mdust/\Mstar$), which is believed to measure the efficiency of dust production or the survival capability of dust grains depending on the literature \citep[e.g.][]{calura_2017:DuststellarMassRatio}. 
We, therefore, first present the resulting dust-to-stellar mass ratio from our galaxies in Fig.~\ref{fig:Mdust_Mstar_relation}, showing both the exact results from our simulations and observations and a schematic representation for ease of reading. We calculate $\Mdust$ and $\Mstar$ for each galaxy run with our fiducial and `enhanced' dust evolution model starting at $z=10$ and stepping $-1$ in redshift up to and including $\zfinal$, with the color denoting the redshift. All values are determined using gas cells and star particles within $0.2\Rvir$ of the galactic center.
We further include two extreme upper bounds on this relation. The first upper bound is determined by the available metal mass in the entire galactic halo of each galaxy. In particular, we determine a maximal dust mass by assuming a constant D/Z $=0.6$, the maximum D/Z predicted by our model and the typical D/Z seen in the densest regions of the MW \citep{jenkins_2009:UnifiedRepresentationGasPhase}, for all gas within $\Rvir$ of the galactic center. We also recalculate $\Mstar$ within $\Rvir$ for this upper bound, but this produces a relatively minor increase.
However, this upper bound is dependent on the fraction of metals and dust ejected from the galactic halo due to the aforementioned bursty star formation. Therefore, the second upper bound is more stringent, representing all metals possibly formed in a galaxy of a given stellar mass assuming a \citet{kroupa_2002:InitialMassFunction} IMF and SNe yields used in FIRE-2 ($M_{\rm metals}\sim0.02\Mstar$).

Comparing with observations, we include results from the ALMA surveys ALPINE \citep{faisst_2020:ALPINEALMAIISurvey,schaerer_2020:ALPINEALMAIISurvey,pozzi_2021:ALPINEALMACIISurvey}, and REBELS \citep{inami_2022:ALMAREBELSSurvey,topping_2022:ALMAREBELSSurvey,sommovigo_2022:ALMAREBELSSurvey,palla_2024:MetalDustEvolution} and various small sample observations from \citet{watson_2015:DustyNormalGalaxy}, \citet{laporte_2017:DustReionizationEra}, \citet{strandet_2017:ISMPropertiesMassive}, \citet{tamura_2019:DetectionFarinfraredIii}, \citet{witstok_2023:EmpiricalStudyDust}, \citet{akins_2023:TwoMassiveCompact}, and \citet{valentino_2024:ColdInterstellarMedium}.
Alternative estimates of $\Mdust$ for the ALPINE sample from \citet{sommovigo_2022:NewLookInfrared} (labelled ALPINE*) are also included.
An approximate $\Mdust$ detection threshold for the REBELS program is also provided similar to \citet{sommovigo_2022:NewLookInfrared}\footnote{Note that for the ALPINE program, assuming a detection threshold of $\sim90\,\mu$Jy (taken from the lowest IR non-detection in \citealt{bethermin_2020:ALPINEALMACIISurvey}), $\Tdust=48 K$, and $\beta=2$ results in a dust mass threshold ${\sim}0.1$ dex higher than REBELS.}.
These observations range in redshift from $z\sim4-8.4$ and are similarly color-coded depending on their redshift. We further include an upper-bound galactic dust mass estimate for GN-z11 at $z=10.6$ from \citet{fudamoto_2024:NOEMAObservationsGNz11}.
Some caution should be taken with the observational estimates of $\Mstar$ and $\Mdust$ provided since they are prone to larger uncertainties than the error bars shown. In regards to $\Mstar$, the assumed SFH can have a large impact on the inferred $\Mstar$. \citet{topping_2022:ALMAREBELSSurvey} found that $\Mstar$ derived with a nonparametric SFH increased by up to 1 dex versus a constant SFH for the REBELS sample. We also highlight that typically a \citet{chabrier_2003:GalacticStellarSubstellar} IMF from $0.1-300 \Msol$ is used (REBELS and ALPINE in particular) compared to the \citet{kroupa_2002:InitialMassFunction} IMF FIRE-2 uses. While the difference in inferred $\Mstar$ is not a simple systematic offset, it is likely on the order of $-0.025$ to $-0.4$ dex \citep{salim_2007:UVStarFormation,pforr_2012:RecoveringGalaxyStellar}.
In regards to $\Mdust$, these estimates typically rely on one ALMA photometric band measurement, and are extremely sensitive to the assumed $\Tdust$. This can be seen with the two ALPINE $\Mdust$ estimates, with the average $\Mdust$ dropping by ${\sim}0.8$ dex between \citet{pozzi_2021:ALPINEALMACIISurvey} and \citet{sommovigo_2022:NewLookInfrared} primarily due to a roughly $2x$ increase in $\Tdust$. 
For now, we leave a detailed discussion on the implications of the uncertainty in $\Tdust$ to Sec.~\ref{sec:observations} and consider both ALPINE and ALPINE* estimates.

Overall, both the fiducial and `enhanced' models predict a roughly power-law $\Mdust/\Mstar$ relation ($\Mdust\propto\Mstar^{\alpha};\, \alpha\sim 1.3$) for $\Mstar\gtrsim10^9\Msol$ irrespective of redshift, with, on average, a ${\sim}0.5$ dex scatter at any given stellar mass. This scatter is due to the bursty star formation of these galaxies, which periodically evacuates large amounts of gas and dust out of the central part of the galaxy. However, the `enhanced' model predicts a relation ${\sim} 0.8$ dex higher than the fiducial model and a larger scatter for $\Mstar\lesssim2\times10^9\Msol$ due to lower $\Zcrit$.
Compared to observations, the fiducial model underpredicts $\Mdust$ across the entire observed $\Mstar$ range, falling $\sim 2 - 0.5$ dex below the REBELS $z\sim7$ observations and even more so for the ALPINE $z\sim5$ observations. 
The `enhanced' model is a better match with observations, overlapping with the $\Mstar\gtrsim 6\times10^9\Msol$ subset of the REBELS and ALPINE*, but still falls 1 dex below the ALPINE observations. 
This large discrepancy with the ALPINE and tentative agreement with ALPINE* suggest that the \citet{pozzi_2021:ALPINEALMACIISurvey} dust masses are overestimated.
This is further bolstered by our upper bound estimates on dust mass, which still fall short of the ALPINE sample, and overlap with the small amount of $\Mstar\gtrsim 2\times10^9\Msol$ galaxies in the REBELS sample. These upper bounds are entirely unrealistic, requiring both maximal dust growth and minimal dust destruction everywhere in the galactic halo. 
Higher, centrally concentrated galactic metallicities in our simulations or a top-heavy IMF could alleviate this tension, but as shown by \citet{marszewski_2024:HighRedshiftGasPhaseMass}, the galaxies from this suite match the observed gas-phase mass-metallicity relation up to the highest redshift where data exists, $z\sim10$.
While a top-heavy IMF cannot be entirely ruled out, as discussed in Sec.~\ref{sec:observations}, based on current metal-budget constraints the ALPINE $\Mdust$ estimates appear overestimated.

Currently, no galactic metallicity measurements exist for the observed high-z DSFGs presented, which has been shown to be the main determinator of efficient dust growth.
In light of JWST's ability to measure galactic metallicity at high-z, we showcase predictions for the relation between galactic dust mass and median D/Z for cool, neutral ($T<1000$ K) gas with respect to the galactic metallicity in Fig.~\ref{fig:dust_Z_relations} for our simulations run with the `enhanced' model.
We define the galaxy-integrated metallicity as the median $\OH$ for gas with $7000<T<15000$ and $\nH>0.5 \,\cmcubed$ to match the properties of nebular regions typically probed by empirical strong emission line methods used in these studies \citep[e.g.][]{nakajima_2023:JWSTCensusMassMetallicity,curti_2024:JADESInsightsLowmass}.
We also account for the depletion of O into dust by only considering gas-phase O instead of total (gas+dust) O abundance and include a $-0.2$ offset to correct for differences in reference O abundances assumed in our simulations \citep{anders_1989:AbundancesElementsMeteoritic} and observations \citep{asplund_2009:ChemicalCompositionSun}. 
We include the handful of galaxies that have metallicity estimates via JWST and accompanying dust mass estimates or upper limits. Notably, GN-z11 \citep{bunker_2023:JADESNIRSpecSpectroscopy,fudamoto_2024:NOEMAObservationsGNz11} and lensed galaxies RXCJ0600-z6.3 \citep{fujimoto_2024:JWSTALMAMultipleline,valentino_2024:ColdInterstellarMedium} and ID4590 \citep{heintz_2023:GasStellarContent,fujimoto_2024:JWSTALMAMultipleline}.
While there are no direct observations of the galactic D/Z versus metallicity relation at $z\gtrsim 5$, we include observations of damped Lyman-$\alpha$ systems (DLAs) out to $z\sim5$ derived from gas-phase element depletions \citep{decia_2016:DustdepletionSequencesDamped, quiret_2016:ESOUVESAdvanced}. 
To highlight some of the underlying uncertainty of this observational method, we show estimates using different [Zn/Fe] calibrations which determine total (gas+dust) metallicity \citep{roman-duval_2022:METALMetalEvolution}. Specifically, we show estimates for the calibration used in \citet{decia_2016:DustdepletionSequencesDamped} and MW-derived calibration from \citet{roman-duval_2022:METALMetalEvolution}.

Our simulations predict $\Mdust$ rises steeply with metallicity, increasing by ${>}2$ dex from $12+\log_{10}({\rm O/H})=7.0-7.8$ with a large ${\sim}1$ dex scatter. 
Above $12+\log_{10}({\rm O/H})\sim7.8$, the slope of $\Mdust$ begins to flatten, and the scatter shrinks. However, only a few galaxies in our suite reach $12+\log_{10}({\rm O/H}) > 8$.
Our predictions agree with the current limited number of observations, however the typical detection threshold of current ALMA programs provides limited constraints on the shape of this relation for low metallicity galaxies.
In regards to D/Z, our simulations predicts a shallow rise with metallicity, with D/Z increasing by ${\sim}1$ dex from $12+\log_{10}({\rm O/H})=7.0-8.4$ with little scatter. There is also a minor increase in the slope at $12+\log_{10}({\rm O/H}) \approx 7.8$, similar to the transition point in $\Mdust$. Our predicted trend matches observations of DLAs using the MW-derived [Zn/Fe] calibration for $12+\log_{10}({\rm O/H}) \lesssim 7.8$ and observations using the \citet{decia_2016:DustdepletionSequencesDamped} calibration for $12+\log_{10}({\rm O/H}) > 7.8$, which could suggest a dependence on the [Zn/Fe] calibration with metallicity. 
We highlight that the ${\rm O/H}$ transition point seen in both trends is close to the $\Zcrit\sim0.05\Zsol$ expected for silicate dust in the `enhanced' model. Therefore, future observations of the $\Mdust$-metallicity relation for low metallicity high-z galaxies would provide further constraints on $\Zcrit$.

\section{Discussion}

As shown in Sec.~\ref{sec:results_observations}, our fiducial simulations cannot reproduce observations of extremely dusty galaxies at $z \gtrsim 5$. Uncertainties in our dust evolution model can explain some of this discrepancy, and differing observational methodologies and biases may explain the rest. We discuss each in detail below and compare our results with other theoretical works.

\subsection{Observational Caveats and Biases} \label{sec:observations}

\subsubsection{Sensitivity to Dust Temperature}

\begin{figure}
    \centering
    \includegraphics[width=0.99\columnwidth]{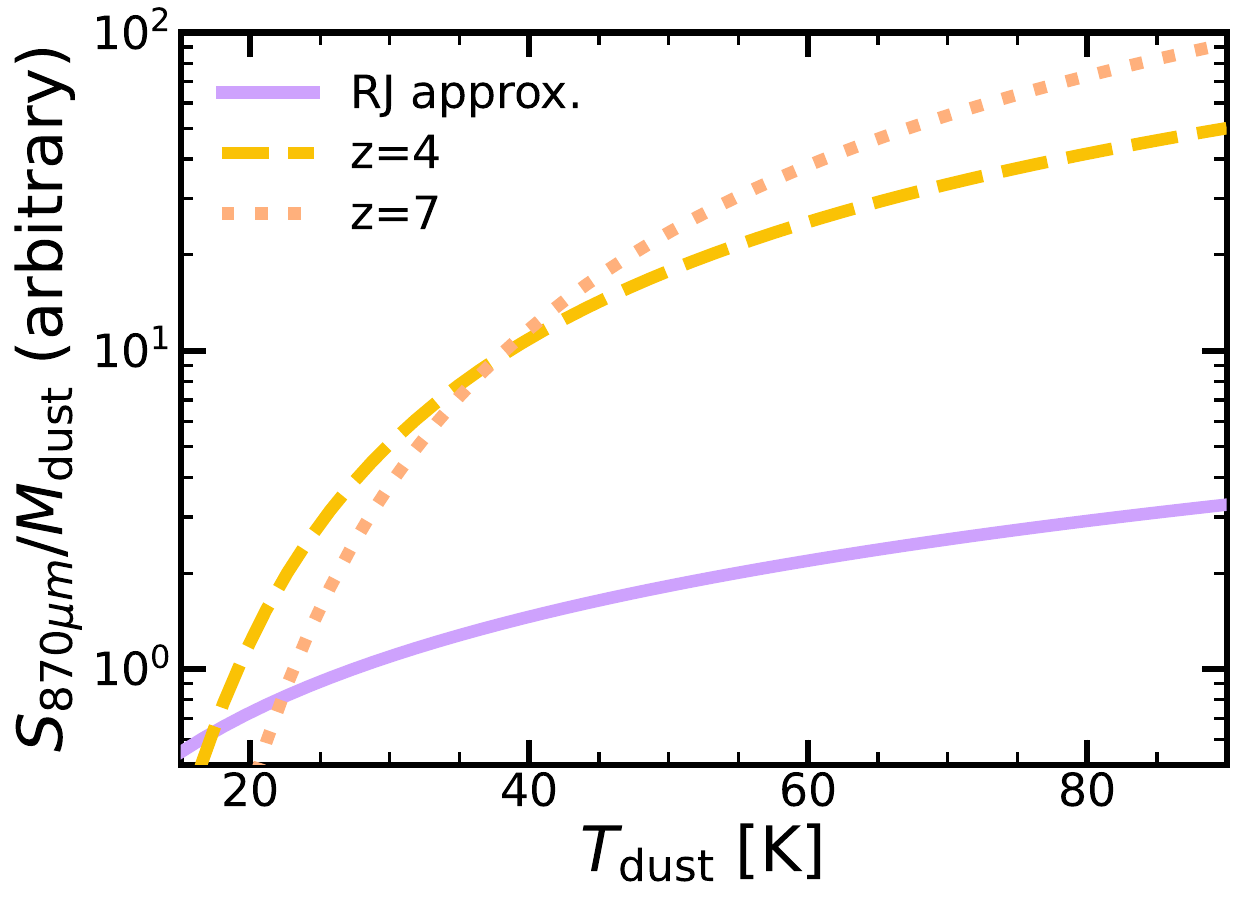}
    \vspace{-0.25cm}
    \caption{Diagram showing the relation between the observed flux density at $870~\micron$ to inferred dust mass ratio $(S_{870\micron}/\Mdust)$ and the assumed dust temperature for set $\beta$ using the RJ approximation ({\it solid}) and not using the RJ approximation for $z=4$ ({\it dashed}) and $z=7$ ({\it dotted}) sources. Note the y-axis units are arbitrary. At high-z, the RJ approximation is not valid, so small changes in $\Tdust$ produce large changes in the inferred $\Mdust$ for a given $S_{870\micron}$.}  
    \label{fig:RJ_diagram}
\end{figure}

Observations of dust emission SEDs at $z\gtrsim5$ are generally limited to one ALMA Band 6 (${\sim}1250\,\micron$) or 7 (${\sim}950\,\micron$) photometric measurement, which probes the rest-frame FIR. 
Dust continuum emission is well described by a modified blackbody (MBB) function \citep[e.g.][]{hildebrand_1983:DeterminationCloudMasses}, and so these observations typically lie at or near the Rayleigh-Jeans (RJ) tail.
Assuming the FIR is optically thin, the relation for the observed flux density at frequency $\nu$ ($S_{\nu}$) is therefore 
\begin{equation}
    S_{\nu}\propto \nu^{\beta} \Mdust B_{\nu}(\Tdust),
\end{equation}
where $\beta$ is the spectral emissivity index determined by dust population properties (grain size and composition), $\Mdust$ is the total dust mass, $B_{\nu}(T)$ is the Planck function, and $\Tdust$ is the effective dust temperature\footnote{The effective dust temperature given by a MBB fitted to the FIR continuum does not correspond to any physical dust temperature. However, simulations suggest it depends on both the luminosity-weighted $\Tdust$, which is dominated by the small amount of hot dust around young stars, and the mass-weighted $\Tdust$, which is dominated by cold dust in the diffuse ISM \citep[][see Fig. 13 for a schematic representation]{liang_2019:DustTemperaturesHighredshift}.}.
Thus, given $\Tdust$ and $\beta$, $\Mdust$ can be determined and vice versa. Since they are difficult to derive observationally, $\Tdust$ and $\beta$ are typically set to assumed values or determined from fitting codes. 
All $\Mdust$ observations presented in Sec.~\ref{sec:results_observations}, besides REBELS sample predictions from \citet{sommovigo_2022:ALMAREBELSSurvey} which we discuss below, utilize one of the above-mentioned methods, with assumed/fitted values in the range of $\Tdust\sim25-60$ K and $\beta\sim1.5-2$. 
These differences in assumed dust properties, specifically $\Tdust$\footnote{We focus on the impacts of $\Tdust$ uncertainties since the uncertainty in $\beta$ appears to have a relatively minor effect on the inferred dust mass in comparison \citep{magnelli_2020:ALMASpectroscopicSurvey,pozzi_2021:ALPINEALMACIISurvey} and does not exhibit a dependence on redshift \citep{witstok_2023:EmpiricalStudyDust,algera_2024:AccurateSimultaneousConstraints}.}, can have a significant impact on the inferred dust mass. 
In particular, the FIR SEDs probed at these redshifts are not entirely in the RJ regime (i.e. cannot approximate $B_{\nu}(\Tdust)\propto\Tdust$), and so relatively small changes to $\Tdust$ can lead to much larger changes to $\Mdust$ \citep{casey_2012:FarinfraredSpectralEnergy,cochrane_2022:DustTemperatureUncertainties}.
We highlight this sensitivity in Fig.~\ref{fig:RJ_diagram}, which shows the relation between the observed flux density to inferred dust mass ratio $(S_{\nu}/\Mdust)$ and the assumed dust temperature for sources at $z=4$ and 7 and using the RJ approximation. We set $\lambda = 870~\micron$ (ALMA Band 7) and assume a constant $\beta$. 

We, therefore, suggest that part of the discrepancy between our simulations and observations is due to too low $\Tdust$ assumptions at these redshifts, which inflate the derived dust mass, and there is observational evidence to support this. In particular, observations of $z\sim5-7$ galaxies with multiple ALMA bands provide constraints on $\Tdust$, but they are more limited in number than single band observations. These multi-band observations generally find these galaxies have high $\Tdust$ \citep[$40 \, {\rm K}\lesssim \Tdust \lesssim 80 \, {\rm K}$;][]{bakx_2020:ALMAUncoversII,faisst_2020:ALMACharacterizesDust, witstok_2023:EmpiricalStudyDust,mitsuhashi_2024:SERENADEIIALMA,valentino_2024:ColdInterstellarMedium,villanueva_2024:ALMACRISTALSurveyDust}, but some galaxies still exhibit low $\Tdust$ \citep[$30 \, {\rm K}\lesssim \Tdust \lesssim 35 \, {\rm K}$;][]{algera_2024:ColdDustLow,algera_2024:AccurateSimultaneousConstraints}.

To showcase the importance of $\Tdust$ estimates, we highlight differing dust mass predictions for the ALPINE sample, the sample our simulations are most in tension with as shown in Sec~\ref{sec:results_observations}. \cite{pozzi_2021:ALPINEALMACIISurvey} made the first dust mass estimates for this sample by assuming a constant $\Tdust=25$ K ($\beta=1.8$), which is a common high-z assumption derived from local observations \citep{scoville_2014:EvolutionInterstellarMedium,scoville_2016:ISMMassesStar,scoville_2017:EvolutionInterstellarMedium}. 
However, more recent estimates by \citet{sommovigo_2022:NewLookInfrared} predict a mean  $\Tdust\sim48$K (with set $\beta=2.0$). This higher $\Tdust$ corresponds to a factor of ${\sim}7$ reduction in the inferred $\Mdust$ compared to \citet{pozzi_2021:ALPINEALMACIISurvey}, putting them in line with the REBELS sample and our simulations. 
These higher $\Tdust$ estimates are determined utilizing a novel method introduced by \citet{sommovigo_2021:DustTemperatureALMA}, which employs the [\textsc{C ii}] 158 $\micron$ line emission as a proxy for $\Mdust$ in order to break the degeneracy between $\Mdust$ and $\Tdust$. 
However we caution that this [\textsc{C ii}] method relies on an assumed dust-to-gas ratio (and with it an assumed metallicity) and nontrivial [\textsc{C ii}]-to-neutral gas mass conversion factor ($\alpha_{\textsc{C ii}}=M_{\rm gas}/L_{\textsc{C ii}}$)\footnote{There is also a small dependence on $\Mstar$. In particular, \citet{sommovigo_2022:NewLookInfrared} recalculated their inferred $\Mdust$/$\Tdust$ relation for the REBELS sample with updated $\Mstar$ values determining by nonparametric SFH fits \citep{topping_2022:ALMAREBELSSurvey}. Despite a ${\lesssim}3\times$ increase in $\Mstar$, $\Mdust$ only slightly increases (see their Fig. 3).}.
\citet{burgarella_2022:ALMAALPINECIISurvey} also predicts higher dust temperatures and lower dust masses for the ALPINE sample ($\Tdust\sim54$ K; $\beta\sim0.87$). Their method uses a composite/stacked IR SED, which incorporates additional MIR Herschel data fitted with a MIR power-law + MBB fit \citep[PL-MBB; ][]{casey_2012:FarinfraredSpectralEnergy} to determine $\Tdust$. However, the effective $\Tdust$ given by a PL-MBB fit is typically higher than that given by a MBB fit due to the sensitivity of the MIR, and total IR luminosity, to the small amount of hot dust near newly formed stars. 
In closing, the intricate details of different dust temperature definitions are beyond the scope of this work, and we point the reader to \citet{liang_2019:DustTemperaturesHighredshift} for an in-depth comparison of these definitions.


\subsubsection{UV-Bright with Detectable IR-Continuum Selection Bias}

Another important aspect to consider is the inherent bias of current high-z observations towards a subset of galaxies that our suite of simulations may not include. The vast majority of observed $z\gtrsim5$ DSFGs are UV-selected galaxies, which biases them towards less-dusty galaxies.
However, these galaxies also have corresponding IR dust continuum measurements, indicating they are on the dustier side. In particular, only ${\sim}20\%$ and ${\sim}40\%$ of the original ALPINE and REBELS UV-selected samples have corresponding IR detections, respectively.
Furthermore, stacking analysis utilizing non-detections for the ALPINE sample further indicates that the median dust masses for the entire sample is $\lesssim0.5$ dex lower than the direct detections \citep{pozzi_2021:ALPINEALMACIISurvey}.
These suggest that current observations are possibly limited to a middle ground in dust content, not being the most dust-rich or the most dust-poor.
However, this could also be a viewing angle effect, with dusty galaxies possibly appearing either UV-bright or dark depending on their orientation \citep{cochrane_2024:DisappearingGalaxiesOrientation}.
We also highlight that, due to their bursty SF, our galaxies spend relatively brief periods in the high SF regime (SFR $\gtrsim10\,\Msol$/yr), as can be seen in Fig.~\ref{fig:enhanced_galaxy_evolution}.
This burstiness affects when these galaxies would be observable in the rest-frame UV \citep[][]{sun_2023:SeenUnseenBursty}, limiting their comparability with observations. In particular, the ALPINE and REBELS samples are limited to SFR $\gtrsim20 \, \Msol$/yr and $\gtrsim40 \, \Msol$/yr, respectively.
Ultimately, more in-depth comparisons with observable quantities will require the creation of mock SEDs using radiative transfer codes, which we leave for future work.

One other important possibility is that observed DSFGs have a metallicity selection bias. 
Our simulations assume a constant \citet{kroupa_2002:InitialMassFunction} IMF, and so would underpredict the metallicity and thus the dust mass, as highlighted in Sec.~\ref{sec:dust_model}, of observed DSFGs if they are dominated by a top-heavy IMF.
Current observations of the stellar mass-metallicity relation, which extends to $z\sim10$, suggest high-z galaxies are metal-poor \citep{nakajima_2023:JWSTCensusMassMetallicity,curti_2024:JADESInsightsLowmass,chemerynska_2024:ExtremeLowmassEnd} and the HiZ FIRE-2 simulation suite matches these observations over the entire observed redshift range
\citep{ma_2016:OriginEvolutionGalaxy,marszewski_2024:HighRedshiftGasPhaseMass} so this possibility seems unlikely.
However, few observations exist within the stellar mass range of observed DSFGs ($\Mstar\gtrsim10^{10}\Msol$), and metallicity estimates for a stacked SED of the ALPINE sample suggest $Z\sim0.5 \, \Zsol$ \citep{vanderhoof_2022:ALPINEALMAIISurvey} which is on the higher end of the observed stellar mass-metallicity relation. Furthermore, one IR-detected $z\sim7$ galaxy with $\Mstar\sim2\times10^9\Msol$ and $Z\sim1.1\,\Zsol$ has been observed \citep{killi_2023:SolarMetallicityGalaxy}, lying well above the observed relation.
Ultimately metallicity estimates for DSFGs are needed to fully rule out this possibility.

\subsection{Dust Model Uncertainties - Fiducial and Enhanced Model Assumptions} \label{sec:dust_model}

While our fiducial dust evolution model reproduces some present-day galaxy-integrated and spatially resolved observations without explicit tuning \citepalias{choban_2024:DustyLocaleEvolution}, there is
${\sim}10$ Gyrs of evolutionary history between $z=5$ and now. This can obfuscate uncertainties in the dust life cycle which can greatly impact $z=5$ predictions but have only minor impacts at present day.
This has led to a vast array of dust evolution models which vary in their assumptions, methodology, and included physics, but all reproduce some present-day observations \citep[e.g.][]{bekki_2015:CosmicEvolutionDust,mckinnon_2016:DustFormationMilky,zhukovska_2016:ModelingDustEvolution,mckinnon_2017:SimulatingDustContent,aoyama_2020:GalaxySimulationEvolution,granato_2021:DustEvolutionZoomcosmological}.
Therefore, it may be one or multiple uncertainties in our model which is the cause of these low high-z dust masses, but would not greatly affect present-day observations. In particular, two uncertainties stand out: 

{\bf (1) Gas-dust accretion timescales:}
Most dust evolution models, including our own, agree that above a `critical' galactic metallicity threshold ($\Zcrit$), dust growth via accretion becomes efficient and the average galactic D/Z rapidly increases 
\citep{hou_2019:DustScalingRelations,li_2019:DustgasDustmetalRatio,graziani_2020:AssemblyDustyGalaxies,parente_2022:DustEvolutionMUPPI}.
They also agree that once accretion becomes efficient, an equilibrium D/Z is eventually reached between dust growth via accretion and dust destruction via SNe shocks. 
However, $\Zcrit$ and the equilibrium D/Z are determined by the average accretion timescale in a galaxy.
If the accretion timescale is increased, then $\Zcrit$ and equilibrium D/Z will increase and decrease, respectively. 
The consequence of this is two-fold.
First, galaxies will need to reach higher metallicities for the onset of efficient accretion, resulting in less time to build sizable dust masses by $z\gtrsim5$. Second, galaxies will have overall lower dust masses at a given metallicity even when accretion is the dominant producer of dust mass.
While the accretion routine in our fiducial model is self-consistent and physically motivated, it is in no sense complete and makes many strong assumptions, such as neglecting dust-gas clumping factors and grain size evolution.
Overall, our routine likely overestimates the gas-dust accretion timescale, as seen when comparing with local universe observations of the galactic D/Z and metallicity relation. In particular, our model predicts a rise in D/Z at ${\sim}0.4$ dex higher metallicities than observed as can be seen in Fig. 8 of \citetalias{choban_2024:DustyLocaleEvolution}.

{\bf (2) SNe II dust yields:} 
As previously mentioned, once accretion becomes efficient, the galactic dust mass increases until an equilibrium between dust growth and destruction is reached. This equilibrium timescale ($\teq$; see Sec 4.2 in \citetalias{choban_2024:DustyLocaleEvolution}) can vary between galaxies, depending on the SNe dust destruction timescale, the fraction of the ISM mass in cold clouds where gas-dust accretion occurs, and the lifetime of those clouds. However, $\teq$ can be universally shortened for all galaxies if the initial amount of dust produced by stars is increased. At $z\gtrsim5$, SNe II are the primary producers of dust, so higher SNe II dust yields could shorten $\teq$. 
In the case of our fiducial model, we assume extremely low SNe II dust yields for all dust species besides carbonaceous dust (see Table~\ref{tab:model_comparison}) taken from \citet{zhukovska_2008:EvolutionInterstellarDust}. These 
creation efficiencies are determined by comparing in situ abundance ratios of presolar dust grains from SNe and AGBs found in meteorites, but these are limited to a handful of grains. Furthermore, both observations \citep[e.g.][]{schneider_2024:FormationCosmicEvolution} and simulations \citep{kirchschlager_2019:DustSurvivalRates,kirchschlager_2023:DustSurvivalRates,kirchschlager_2024:TotalDestructionComplete} predict both extremely high and extremely low SNe II dust production.

To test these uncertainties in our fiducial model, we reran simulations of {\bf z5m12a} with various reasonable changes (i.e. within theoretical/observational uncertainty) to our accretion and SNe dust production routines, which are presented in Appendix~\ref{app:dust_evo_variations}. 
In summary, we find that both decreasing the accretion timescale by a factor of 4 and increasing the SNe II dust creation efficiency to 20\% for carbonaceous, silicates, and metallic iron (set to 15\%, 0.035\%, and 0.1\% respectively in the fiducial model) are likely needed, which is the basis for the `enhanced' model. We also provide a direct quantitative comparison of SNe dust mass yields and accretion timescales in Table~\ref{tab:model_comparison}. The decreased accretion timescale allows for efficient accretion to commence well before $z=7$ and results in a ${\sim}1$ dex higher equilibrium D/Z. The additional increase in the SNe creation efficiency increases D/Z by ${\sim}1$ dex at early times and decreases the time it takes to reach an equilibrium D/Z.
As shown in Fig.~\ref{fig:Mdust_evolution}, this results in an average ${\lesssim}1$ dex increase in the galactic dust mass for our galaxies at $z\sim5-7$. 
In the end, however, we are mainly limited by the metal budget of our galaxies. All of our galaxies have median $Z\lesssim0.4\Zsol$ at $\zfinal$, and so even assuming the most extreme case of D/Z $=0.6$ everywhere in the halo, we cannot reproduce a large number of $\Mdust$ observations as shown in Fig.~\ref{fig:Mdust_Mstar_relation}. A top-heavy IMF could alleviate this issue, but the HiZ FIRE-2 simulation suite currently matches the observed stellar mass-metallicity relation up to the highest redshift where data exists, $z\sim10$ \citep{ma_2016:OriginEvolutionGalaxy,marszewski_2024:HighRedshiftGasPhaseMass}. 
Therefore, unless observed DSFGs do not follow the observed stellar mass-metallicity relation, it seems unlikely that our galaxy evolution model is to blame. This gives us further credence that many high-z $\Mdust$ measurements are overpredicted, as discussed in Sec.~\ref{sec:observations}.

We highlight that the above changes would only marginally affect the local universe results presented in \citetalias{choban_2024:DustyLocaleEvolution}. In particular, the increased SNe II dust creation efficiencies would result in slightly higher D/Z for low-metallicity galaxies, which are dominated by SNe and AGB dust production, and shorten the predicted $\teq$ by ${<}50\%$. The reported $\Zcrit$'s for each dust species would also decrease by roughly the same factor as their accretion timescales. This could cause issues for \citetalias{choban_2024:DustyLocaleEvolution} finding that the high $\Zcrit$ for carbonaceous dust explains the lack of small carbonaceous grains in low-metallicity galaxies.
However, we stress that only the accretion timescale for silicate dust needs to be shorter since silicates represent the majority of the maximum formable amount of dust.

\subsection{Previous Theoretical Works} \label{sec:simulation_comparison}

Here, we compare our simulation predictions to the theoretical field at large and provide insights for assumed high-z dust populations used in simulation post-processing to produce observable quantities.

\subsubsection{Comparisons to Other High-z Dust Evolution Models}

In recent years, an increasing number of works utilizing dust evolution models integrated into semi-analytical models and galaxy simulations have focused on explaining local universe dust observations. However, relatively few of these have been turned towards $z\gtrsim5$ observations.
The current high-z galaxy simulation landscape includes 
\citet{graziani_2020:AssemblyDustyGalaxies} and \citet{dicesare_2023:AssemblyDustyGalaxies} cosmological simulations run with {\small DUSTYGADGET},
\citet{esmerian_2022:ModelingDustProduction,esmerian_2024:ModelingDustProduction} post-processed simulations from the Cosmic Reionization on Computers project \citep{gnedin_2014:CosmicReionizationComputers}, \citet{lewis_2023:DUSTiERDUSTEpoch} DUSTiER cosmological simulation run with RAMSES-CUDATON \citep{ocvirk_2016:CosmicDawnCoDa},  and \citet{lower_2023:CosmicSandsOrigin,lower_2024:CosmicSandsII} Cosmic Sands cosmological zoom-in simulations run with SIMBA \citep{dave_2019:SimbaCosmologicalSimulations}.
Despite the differing galaxy formation models used, all of these works predict that gas-dust accretion is the dominant producer of dust mass for galaxies in the observed stellar mass range ($\Mstar\gtrsim10^9\Msol$), similar to our findings. However, they all produce a similar dust-to-stellar mass relation, with relatively little scatter within each study, that is ${\gtrsim}0.5$ dex higher than our `enhanced' model predictions, and tentatively agree\footnote{For clarity, \citet{esmerian_2024:ModelingDustProduction} only includes galaxies with $\Mstar\lesssim10^9\Msol$, but these agree with the handful of galaxies at the lower stellar mass end of the ALPINE and REBELS samples (see their Fig. 4). \citet{lower_2023:CosmicSandsOrigin,lower_2024:CosmicSandsII} only compare their simulations at $z=6.7$ with the REBELS sample, roughly agreeing with the entire sample (see their Fig. 7).} with both low stellar mass ALPINE and most REBELS dust masses.

The cause of this discrepancy with our results is likely due to the lower resolution of previous works (${\gtrsim} 2 $ dex lower than our own). These simulations do not resolve the multi-phase ISM, instead relying on sub-resolution prescriptions that result in the overprediction of $\Mdust$ at high-z as we describe below.
Firstly, these works employ sub-grid star formation schemes, which result in systematically less bursty star formation histories \citep[i.e.][]{iyer_2020:DiversityVariabilityStar}.
Case in point, the Cosmic Sands simulations, the highest resolution simulations utilized in previous works, predict an effectively monotonic increasing SFH \citep[][see their Fig. 3]{lower_2023:CosmicSandsOrigin}.
This lack of bursty star formation means these simulations do not experience strong blowouts of their cold gas reservoir as indicated by our simulations, and so dust growth via accretion is likely more efficient on average. Furthermore, this can suppress any predicted scatter in the stellar-to-dust mass relation.
Secondly, these works employ a one-phase gas-dust accretion routine. This means accretion occurs in effectively all gas phases \citep[e.g.][]{lewis_2023:DUSTiERDUSTEpoch,lower_2023:CosmicSandsOrigin,esmerian_2022:ModelingDustProduction}, instead of being restricted to cold gas where it must then be cycled out to warmer phases. Furthermore, numerous `sub-grid' dust and gas physical processes are grouped into an overall accretion timescale normalization factor which is resolution and implementation dependent\footnote{\citet{lewis_2023:DUSTiERDUSTEpoch} in particular has extremely short accretion timescales since they do not scale with the local metallicity. This leads to accretion being so efficient that all galaxies with $\Mstar\gtrsim10^6\Msol$ have the maximum amount of dust possible.}.
As shown in \citetalias{choban_2022:GalacticDustmodellingDust}, when in the accretion-dominated regime, such a routine results in a relatively constant D/Z for all but the hottest gas phase within a MW-like galaxy.
The combined effect of these two assumptions is an increased D/Z in all gas phases to the point where many of these simulations reach their maximum D/Z $\gtrsim40\%$ for most gas in the galactic halo \citep{graziani_2020:AssemblyDustyGalaxies,dicesare_2023:AssemblyDustyGalaxies,lewis_2023:DUSTiERDUSTEpoch}.

In contrast, works utilizing semi-analytical models have more mixed results. 
Notably, the DELPHI model used by \citet{dayal_2022:ALMAREBELSSurvey} and \citet{mauerhofer_2023:DustEnrichmentEarly} and the CHEMEVOL model used by \citet{palla_2024:MetalDustEvolution} predict that accretion is subdominant to dust creation by SNe II at these epochs. Meanwhile, the \citet{popping_2017:DustContentGalaxies} model, \citet{vijayan_2019:DetailedDustModelling}, and L-GALAXIES model  \citet{triani_2020:OriginDustGalaxies} Dusty SAGE model find accretion is the dominant producer of dust mass. 
Despite these differences, all of these works underpredict dust masses relative to some observations similar to our results, such as the lower stellar mass end of the REBELS samples and \citet{pozzi_2021:ALPINEALMACIISurvey} ALPINE estimates.
Furthermore, some works conducted a `maximal dust model' {\it gedankenexperiment} based on the available metal budget, similar to our $\Mdust$ upper bound, and find many observations fall on or above this maximum limit \citep{vijayan_2019:DetailedDustModelling,dayal_2022:ALMAREBELSSurvey}.

\subsubsection{Insights for Post-Processing Observables}

Due to the numerous observational uncertainties at high-z, as we discuss in Sec.~\ref{sec:observations}, and upcoming observations with JWST, there has been a recent surge in post-processing existing simulations that do not include dust evolution, including the FIRE simulations, to predict observables. 
This process utilizes radiative transfer codes coupled with assumed dust populations to create mock-SEDs from which direct observational predictions can be made 
\citep{liang_2018:SubmillimetreFluxProbe,liang_2019:DustTemperaturesHighredshift,cochrane_2019:PredictionsSpatialDistribution,cochrane_2022:DustTemperatureUncertainties,cochrane_2023:PredictingSubmillimetreFlux,cochrane_2024:DisappearingGalaxiesOrientation,ma_2019:DustAttenuationDust,vogelsberger_2020:HighredshiftJWSTPredictions,shen_2020:HighredshiftJWSTPredictions,parsotan_2021:RealisticMockObservations,pallottini_2022:SurveyHighzGalaxies,shen_2022:HighredshiftPredictionsIllustrisTNG,vijayan_2022:FirstLightReionisation,katz_2023:SPHINXPublicData}.
The typical dust populations assumed are quite simplistic, with the standard being a constant D/Z for all gas within the galactic halo below a set temperature above which dust is presumed to be instantly destroyed by sublimation/sputtering (typically D/Z${\sim}0.4$; $T{\lesssim}10^6$ K).
However, as we show in Fig.~\ref{fig:enhanced_dust_evolution}, high-z galaxies likely have lower D/Z on average, along with a large variation between gas phases, which can have a large impact on observables.

For example, current theoretical works that attempt to constrain mass-weighted $\Tdust$ in the early universe vary considerably due to differing dust population assumptions. 
Post-processed $z\gtrsim2$ FIRE simulations, which utilized an assumed dust population, suggest $\Tdust\sim25\pm7$ K with little evolution over redshift \citep{ma_2019:DustAttenuationDust,liang_2019:DustTemperaturesHighredshift}. 
Meanwhile, post-processed SIMBA simulations that use self-consistently evolved dust populations predict higher dust temperatures on average along with a large variation between galaxies ($\Tdust\sim30-80$ K; \citealt{lower_2024:CosmicSandsII}). However, the SIMBA simulations have ${>}2$ dex lower resolution and do not resolve the multi-phase ISM, so this is not a one-to-one comparison.
Therefore, we advise that future post-processing works utilize realistic dust populations derived from simulations like those presented here. We also plan to investigate the sensitivity of post-processed predictions in future works.

\section{Conclusions} \label{sec:conclusions}

In this work, we investigate the evolution and buildup of dust in moderately massive galaxies at $z\gtrsim5$ utilizing a suite of 8 cosmological zoom-in simulations, originally simulated in \citet{ma_2018:SimulatingGalaxiesReionization,ma_2019:DustAttenuationDust} with the FIRE-2 model \citep{hopkins_2018:FIRE2SimulationsPhysics} for stellar feedback and ISM physics, rerun with the ``Species'' dust evolution model \citepalias{choban_2022:GalacticDustmodellingDust}. This dust evolution model accounts for dust creation in stellar outflows (SNe II/Ia and AGBs), growth from gas-phase accretion, destruction from SNe shocks, thermal sputtering, and astration, and turbulent dust and metal diffusion in gas. It tracks the evolution of specific dust species (silicates, carbon, silicon carbide), treating each uniquely depending on their chemical composition, along with theoretical nano-particle metallic iron (Nano-iron) dust species and an oxygen-bearing (O-reservoir) dust species. It also incorporates a physically motivated dust growth routine, which accounts for Coulomb enhancement and CO formation in dense molecular environments. 
Due to these details, this model is able to replicate a wide range of present-day observations \citepalias{choban_2024:DustyLocaleEvolution}.

The 8 galaxies we selected cover a range of stellar masses ($10^9 \, \Msol \leq M_{*} \leq 10^{10} \, \Msol$) and final redshifts $\zfinal\sim10-5$ (Fig.~\ref{fig:mock_JWST1} and~\ref{fig:mock_JWST2}) similar to dusty, star-forming galaxies observed by the ALMA ALPINE ($z\sim5$) and REBELS ($z\sim7$) surveys. We summarize our findings on high-z dust buildup below:

\begin{enumerate}
    \item Accretion is the dominant producer of dust mass for these galaxies. However, our fiducial model cannot reproduce observed dust masses at any redshift or stellar mass (Fig.~\ref{fig:Mdust_Mstar_relation}) due to two factors. 
    First, our simulations predict these galaxies have low metallicity ($Z<0.5 \, \Zsol$) and extremely bursty star formation
    (Fig.~\ref{fig:enhanced_galaxy_evolution}). These attributes delay the onset of efficient dust growth via accretion, which is determined by a critical metallicity threshold ($\Zcrit$), and limit the efficiency of accretion, due to the continuous disruption of cold gas where accretion occurs. 
    Second, our model assumes low SNe II dust creation efficiencies (Table~\ref{tab:model_comparison}), resulting in extremely low initial D/Z $<0.01$. This prolongs the buildup of dust mass once accretion becomes efficient (Appendix~\ref{app:dust_evo_variations}).

    \item Our `enhanced' model incorporates modest changes that are within theoretical/observational uncertainty (i.e. decreasing both the accretion timescale by a factor of 4 and increasing SNe II creation efficiencies to 20\%; see Table~\ref{tab:model_comparison}). These changes result in a ${\sim}1$ dex increase in both the initial D/Z and the $\Mdust$ at $\zfinal$ (Fig.~\ref{fig:Mdust_evolution} and~\ref{fig:enhanced_dust_evolution}). This agrees with observed $\Mdust$ estimates for $\Mstar>6\times10^9\Msol$ from the REBELS survey but still falls ${\sim}1$ dex below \citet{pozzi_2021:ALPINEALMACIISurvey} estimates for the ALPINE survey.
    These results suggest SNe II are efficient dust producers and highlight that our accretion routine is likely missing an important physical process.

    \item Given the low predicted metallicities of our simulated galaxies, the \citet{pozzi_2021:ALPINEALMACIISurvey} $\Mdust$ estimates for the ALPINE survey may be overestimated (Fig.~\ref{fig:Mdust_Mstar_relation}), likely due to low $\Tdust\sim25\,K$ assumptions (Fig.~\ref{fig:RJ_diagram}). 
    Follow-up multi-band IR observations generally find $40 \, {\rm K}\lesssim \Tdust \lesssim 80 \, {\rm K}$, and alternative ALPINE $\Mdust$ estimates from 
    \citet{sommovigo_2022:NewLookInfrared}, which predict $\Tdust\sim50\,K$, tentatively support this. Future post-processing of our simulations with radiative transfer codes is also needed to confirm this.

    \item Future observations to determine the burstiness of high-z galaxies are needed to constrain the disruption of cold gas and thus the overall efficiency of gas-dust accretion in the early Universe.
    Previous high-z cosmological simulations incorporating dust evolution models rely on sub-resolution prescriptions for star formation and gas-dust accretion routines resulting in a higher accretion efficiency. In particular, they produce a less bursty star formation history, allowing for continuous gas-dust accretion, and typically do not restrict accretion to cold gas, allowing accretion to occur everywhere in a galaxy.

    \item Works utilizing radiative transfer codes to post-process simulations typically assume a dust population with a constant D/Z, but a spatially variable D/Z should be considered to understand its effects on observables. In particular, we find a maximum D/Z $\sim0.3$ in cool ($T<1000$ K) gas which decreases by ${\lesssim}1$ dex for hot ($T\geq10^4$ K) gas in these high-z galaxies (Fig.~\ref{fig:enhanced_dust_evolution}).

    \item JWST allows for metallicity measurements of these high-z DSFGs, which is the primary determinator of dust evolution. We, therefore, provide predictions for the relation between galactic dust mass and D/Z with galactic metallicity (Fig.~\ref{fig:dust_Z_relations}). Ultimately, observations of low-metallicity galaxies with lower ALMA detection thresholds than current programs are needed to better constrain $\Zcrit$.

\end{enumerate}

\section*{Acknowledgements}
We thank Denis Burgarella for sharing details of his work, Laura Sommovigo, Guochao Sun, and Robert Feldmann for insightful discussions, and 
Alejandro Guzmán-Ortega
and Sanchit Sabhlok for their insights into creating composite images.
DK was supported by NSF grant AST-2108324. 
This research was supported in part by Lilly Endowment, Inc., through its support for the Indiana University Pervasive Technology Institute.
The authors acknowledge the Indiana University Pervasive Technology Institute \citep{stewart_2017:IndianaUniversityPervasive} for providing supercomputing, database, and storage resources that have contributed to the research results reported within this paper.
We ran simulations using: the Extreme Science and Engineering Discovery Environment
(XSEDE), supported by NSF grant ACI-1548562; Frontera allocations AST21010 and AST20016, supported by the NSF and TACC; Big Red 200 at the Indiana University Pervasive Technology Institute.
The data used in this work were, in part, hosted on facilities supported by the Scientific Computing Core at the Flatiron Institute, a division of the Simons Foundation. This work also made use of MATPLOTLIB \citep{hunter_2007:Matplotlib2DGraphics}, NUMPY \citep{harris_2020:ArrayProgrammingNumPy}, SCIPY \citep{virtanen_2020:SciPy10Fundamental}, and NASA’s Astrophysics Data System.

\datastatement{The data supporting the plots within this article are available on reasonable request to the corresponding author. A public version of the GIZMO code is available at \gizmourl.}



\bibliographystyle{mnras}
\bibliography{references} 



\appendix

\section{Effects of Dust Model Variations and ISM Evolution} \label{app:dust_evo_variations}

\begin{table*}
        \renewcommand{\arraystretch}{1.15}
	\centering
	\begin{tabular}{l c c c c c c c c m{30mm}} 
		\hline
		Name & $M_{*}$ & $R_{\rm *, 1/2}$ & $M_{\rm gas, neutral}$ & $R_{\rm neutral, 1/2}$ & Z & $M_{\rm dust}$ & $R_{\rm dust, 1/2}$ & SFR$_{\rm 10\,Myr}$ & Notes \\
             & $(\Msol)$ & (kpc) & $(\Msol)$ & (kpc) & $(\Zsol)$ & $(\Msol)$ & (kpc) & $(\Msol/{\rm yr})$ & \\
		\hline
            z5m12a & $5.43 \times 10^{9}$ & 6.23 & $1.91 \times 10^{10}$ & 5.43 & 0.12 &  $1.08 \times 10^{6}$ & 6.19 & 5.15 & fiducial dust model\\
            z5m12a\_4acc &  $5.59 \times 10^{9}$ & 3.78 & $2.41 \times 10^{10}$ & 4.69 & 0.10 &  $5.99 \times 10^{6}$ & 3.78 & 6.59 & 1/4$\times$ accretion timescale \\
            z5m12a\_0.9sd &  $5.39 \times 10^{9}$ & 3.83 & $2.73 \times 10^{10}$ & 4.75 & 0.11 &  $6.83 \times 10^{6}$ & 4.49 & 35.28 & 90\% SNe II stardust  \\
            z5m12a\_enh &  $4.92 \times 10^{9}$ & 5.37 & $2.27 \times 10^{10}$ & 5.11 & 0.13 &  $7.08 \times 10^{6}$ & 4.79 & 7.60 &  20\% SNe II stardust \& \newline 1/4$\times$ accretion timescale \\
            z5m12a\_enh\_Tcut &  $4.55 \times 10^{9}$ & 4.66 & $2.40 \times 10^{10}$ & 3.59 & 0.09 &  $8.55 \times 10^{6}$ & 2.46 & 14.64 & 20\% SNe stardust \& \newline 1/4$\times$ accretion timescale \& \newline T$_{\rm cut} = 1000 K$\\
            z5m12a\_enh\_FIRE3 &  $6.94 \times 10^{9}$ & 1.98 & $1.54 \times 10^{10}$ & 4.17 & 0.18 &  $2.38 \times 10^{6}$ & 2.45 & 30.21 & 20\% SNe stardust \& \newline 1/4$\times$ accretion timescale \& \newline FIRE-3 physics\\
	   \hline
	\hline
	\end{tabular}
	\caption{Parameters describing galactic properties at simulation end for reruns of {\bf z5m12a} with various changes to our dust evolution model and/or with FIRE-2 vs FIRE-3 ISM physics and feedback model. Columns {\bf (1)}-{\bf (9)} are the same as columns {\bf (1)},{\bf (5)}-{\bf (12)} in Table~\ref{tab:simulations}, with one additional column {\bf (10)} Notes for each simulation on the specific changes made to the model. Overall, the final galactic properties, besides dust mass, change little between reruns with FIRE-2. FIRE-3 produces higher stellar mass and lower metal mass due to enhanced star formation at early times and updated stellar yields, which are more in line with \citet{asplund_2009:ChemicalCompositionSun} abundances.}
    \label{tab:z5m12a_simulations}
\end{table*}

To test how uncertainties in our dust evolution model affect the resulting dust population evolution and final dust mass, we reran simulations of one galaxy in our suite ({\bf z5m12a}) with mostly reasonable variations to our dust evolution model. In particular, we tested {\bf (1)} decreasing the accretion timescale by a factor of 4, {\bf (2)} increasing the SNe II dust creation efficiency of each dust species to 90\%, {\bf (3)} same as (1) and increasing the SNe II dust creation efficiency of each dust species to 20\% (labeled the `enhanced' model in the main text), and {\bf (4)} same as (3) and increasing the temperature cutoff of gas-dust accretion ($T_{\rm cut}=1000$ K). 
We also test the robustness of our model to variations in ISM evolution by running one simulation with both the `enhanced' dust evolution model and the FIRE-3 version \citep{hopkins_2023:FIRE3UpdatedStellar} of the FIRE code.
FIRE-3 makes a variety of improvements to the stellar inputs and numerical methods, focusing in particular on updating the stellar evolution tracks used for stellar feedback and nucleosynthesis with newer, more detailed models, as well as improving the detailed thermochemistry of cold atomic and molecular gas, and adopting the newer \citet{asplund_2009:ChemicalCompositionSun} proto-solar reference abundances with $\Zsol\sim0.014$.
The resulting galactic properties at $\zfinal$ for each of these reruns are provided in Table~\ref{tab:z5m12a_simulations}. 

\begin{figure*}
    \centering
    \plotsidesize{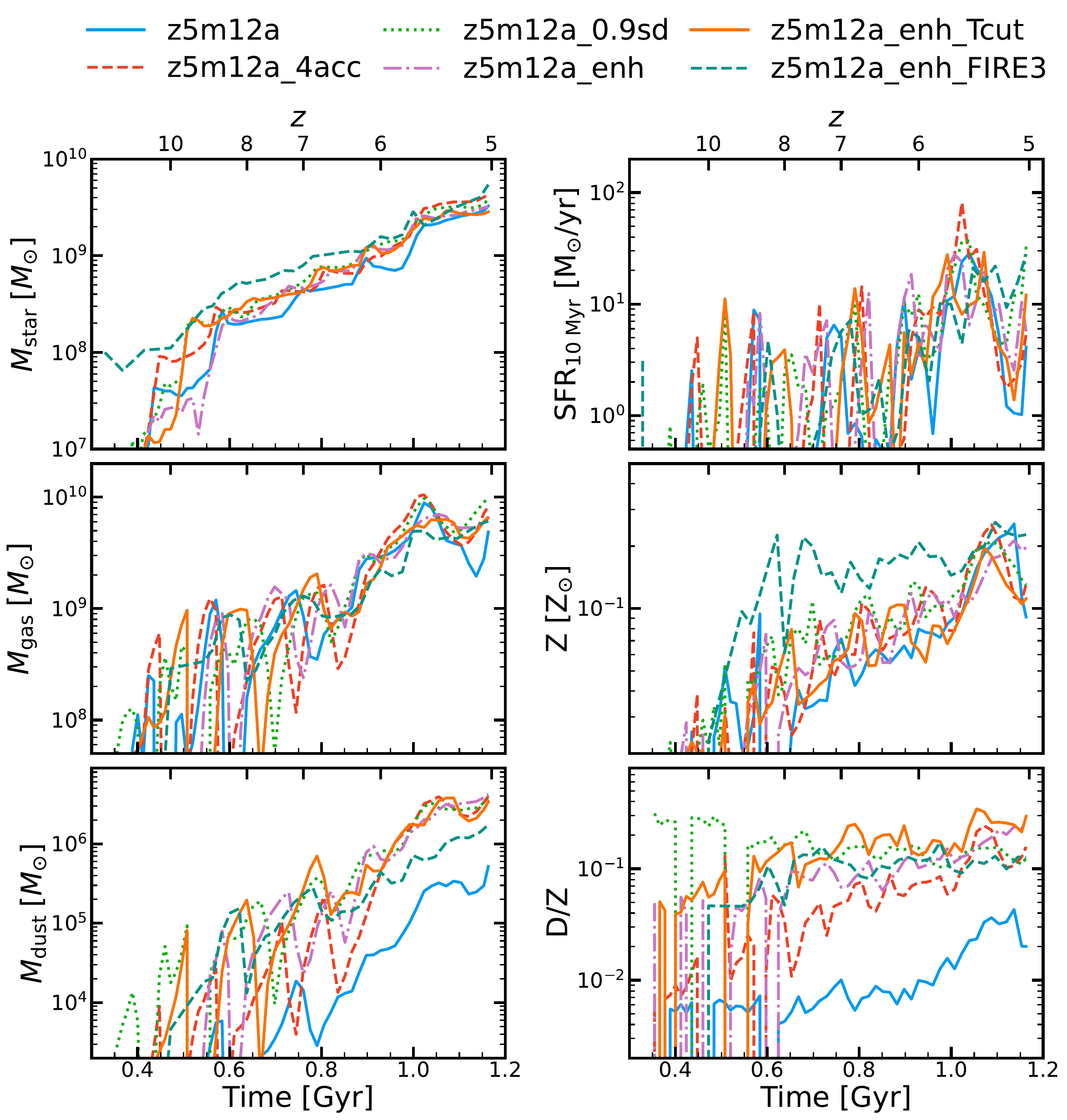}{0.9}
    \vspace{-0.25cm}
    \caption{Evolution of total stellar mass ({\it top left}), star formation averaged over 10 Myr intervals ({\it top right}), gas mass ({\it middle left}), median metallicity in cool ($T<1000$ K) gas ({\it middle right}), dust mass ({\it bottom left}), and median D/Z of cool gas ({\it bottom right}) within 0.2$\Rvir$ for reruns of {\bf z5m12a} with the varying modifications to our dust evolution model along with one simulation rerun with the FIRE-3 ISM physics and feedback model. We note that despite stochastic variations between FIRE-2 reruns, the resulting galactic evolution is largely unchanged. On the other hand, FIRE-3 has a higher initial SFR, resulting in a higher metallicity at early times, but produces similar galactic properties at $\zfinal$ as FIRE-2. Various minor changes to our gas-dust accretion routine and SNe II dust creation efficiencies can increase the final galactic dust mass by ${\sim}1$ dex. Compared to FIRE-2, FIRE-3 produces a lower galactic dust mass primarily due to lower amounts of cold, dense gas due to its updated star formation criteria.}
    \label{fig:z5m12a_galaxy_evo}
\end{figure*}

\begin{landscape}
\begin{figure}
    \centering
    \includegraphics[width=0.95\columnwidth]{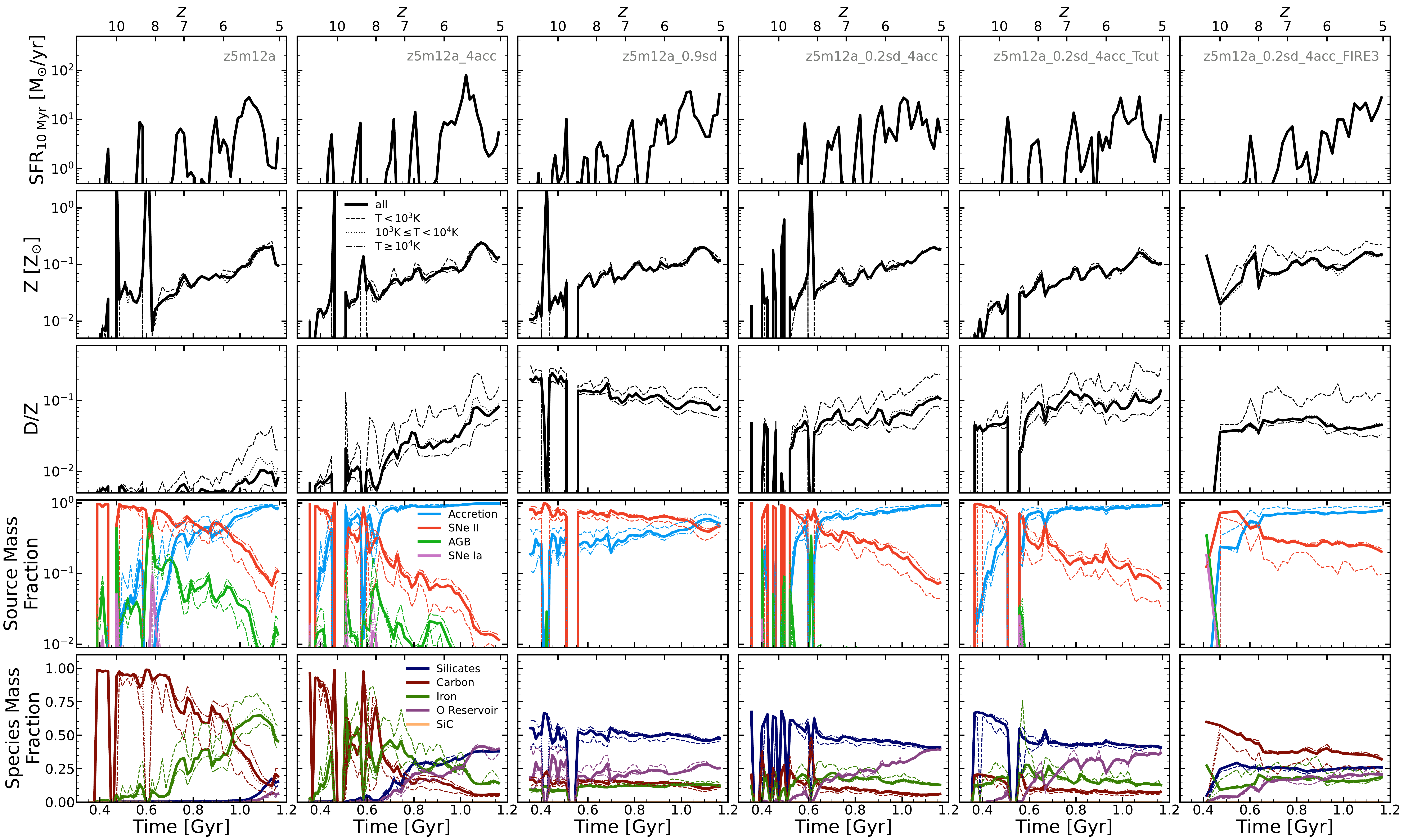}
    \vspace{-0.25cm}
    \caption{Same as Fig.~\ref{fig:enhanced_galaxy_evolution} for reruns of {\bf z5m12a} with variations to our dust evolution model and/or with the FIRE-3 ISM physics and feedback model as specified in Table~\ref{tab:z5m12a_simulations}. Decreasing the accretion timescale alone results in an earlier onset of efficient accretion and ${\sim} 1$ dex higher D/Z at $\zfinal$. However, when this is coupled with low initial D/Z, due to low SNe II dust creation efficiencies, the buildup of dust mass is slow. Increasing SNe II creation efficiencies produce initially higher D/Z, but if this is not coupled with faster accretion D/Z drops over time due to the destruction of dust by SNe shocks. Increasing the temperature cutoff for gas-dust accretion results in a marginally faster increase in D/Z and a slightly higher median D/Z. FIRE-3 predicts a higher initial carbonaceous dust mass, due to updated SNe II yields, and a lower D/Z, due to a systematic reduction in the amount of cool, dense gas.}
    \label{fig:z5m12a_dust_evo}
\end{figure}
\end{landscape}

Fig.~\ref{fig:z5m12a_galaxy_evo} shows the evolution of various galactic properties for each rerun, specifically the stellar mass, star formation rate average over 10 Myr, gas mass, median metallicity of cool ($T<1000$ K) gas, dust mass, and median D/Z of cool gas. In
Fig.~\ref{fig:z5m12a_dust_evo} we compare a detailed breakdown of each simulation's metal and dust population evolution. 
Focusing first on galactic properties, it can be seen that despite stochastic variations between each rerun with FIRE-2, there is little variation in the overall galactic evolution besides the exact timing of star formation bursts. The rerun with FIRE-3 exhibits higher star formation rates at early times, similar to what is seen in \citet{hopkins_2023:FIRE3UpdatedStellar}, resulting in a higher stellar mass and metallicity early on, but roughly matches FIRE-2 galactic properties by $\zfinal$.

Looking at each galaxy's dust population evolution, we can see modest variations in the dust evolution model result in considerable evolutionary changes and dust masses at $\zfinal$. Decreasing the accretion timescale results in an earlier onset of efficient accretion (i.e. lower $\Zcrit$) and a higher overall dust mass and median D/Z. However, when this is coupled with low SNe dust creation efficiency, as with model (1), the timescale of dust buildup via accretion can be long (${\gtrsim}0.4$ Gyr), which may be too long to produce $z\sim7$ DSFGs. Increasing SNe II creation efficiency alone, and to its most extreme case as with model (2), results in an initially high median D/Z $\sim0.2$, but this decreases over time as more dust is destroyed by SNe shocks then can be created by SNe II and accretion. 
This highlights the difficulty of producing $z\sim5$ DSFGs with primarily SNe-created dust, even when considering near maximum SNe dust creation.
Modest decreases and increases to the accretion timescale and SNe II dust creation efficiency respectively, as with model (3), result in both an earlier increase in D/Z, due to efficient accretion, and a faster increase in D/Z, due to the initially higher D/Z produced by SNe dust creation.
Further increasing the temperature cutoff for gas-dust accretion, as with model (4). has little effect on the overall evolution and only slightly increases the median D/Z.

Comparing the predicted dust evolution from FIRE-2 and FIRE-3 with model (3), FIRE-3 predicts an initially carbonaceous-dominated dust population due to its updated stellar yields (from the synthesis of \citealt{nomoto_2013:NucleosynthesisStarsChemical,pignatari_2016:NuGridStellarData,sukhbold_2016:CorecollapseSupernovae9, limongi_2018:PresupernovaEvolutionExplosive,prantzos_2018:ChemicalEvolutionRotating}), which have higher C and lower Si and Fe SNe II yields.
Despite producing a higher metallicity at earlier times, FIRE-3 predicts a rise in D/Z via accretion at roughly the same time as FIRE-2 due to lower Si and Fe abundances at a given metallicity (\citet{anders_1989:AbundancesElementsMeteoritic} vs \citet{asplund_2009:ChemicalCompositionSun}). 
FIRE-3 also predicts a lower median D/Z due to an overall decrease in cold gas compared to FIRE-2. In particular, FIRE-3 predicts ${\sim}2\times$ less gas mass with $T<1000$ K compared to FIRE-2.

\section{Additional Figures} \label{app:addtional_figs}

\begin{figure*}
    \centering
    \plotsidesize{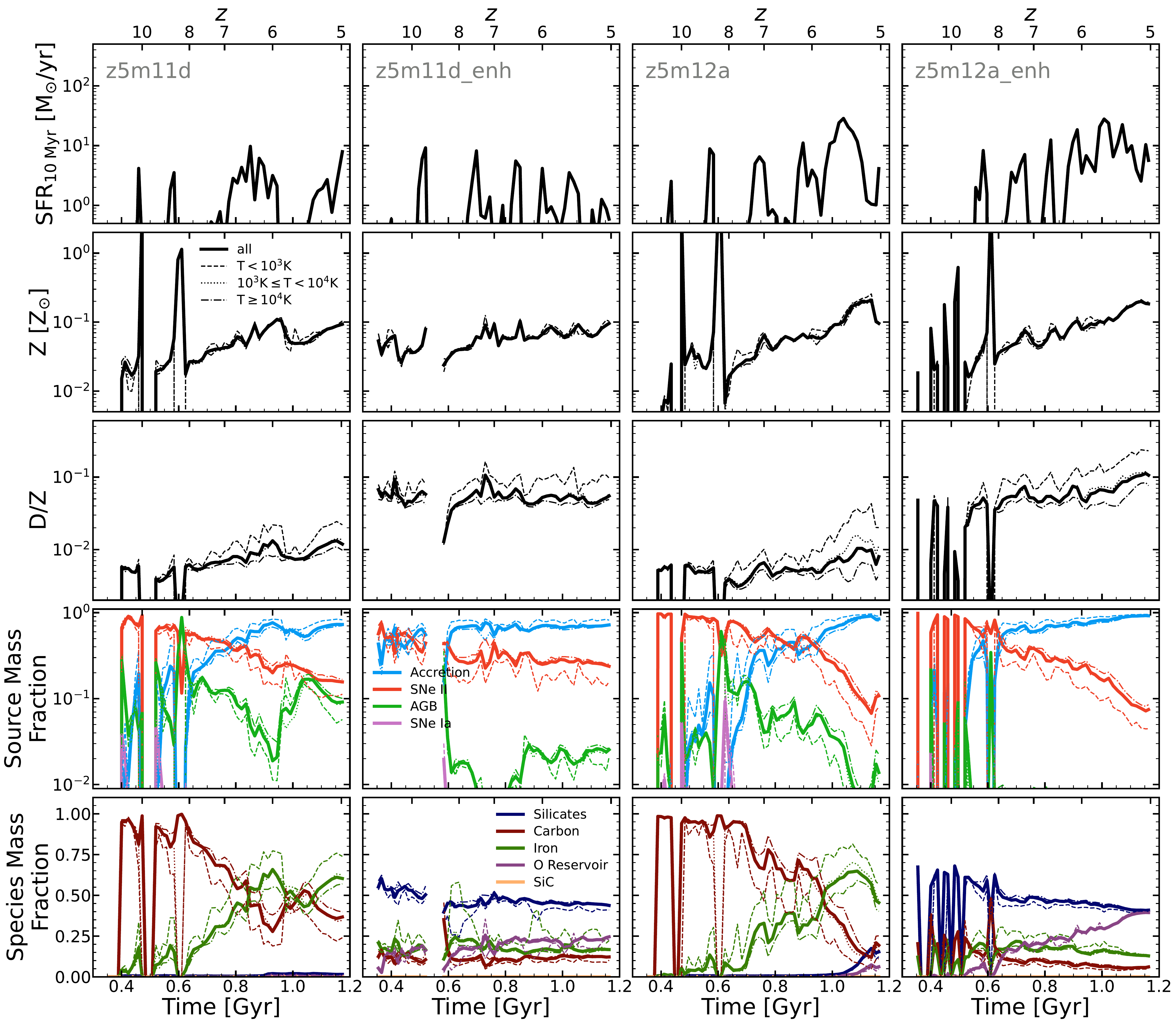}{0.9}
    \vspace{-0.25cm}
    \caption{Same as Fig.~\ref{fig:enhanced_dust_evolution} for {\bf z5m11d} and {\bf z5m12a}.}
    \label{fig:app_dust_evo_comparison1}
\end{figure*}

\begin{figure*}
    \centering
    \plotsidesize{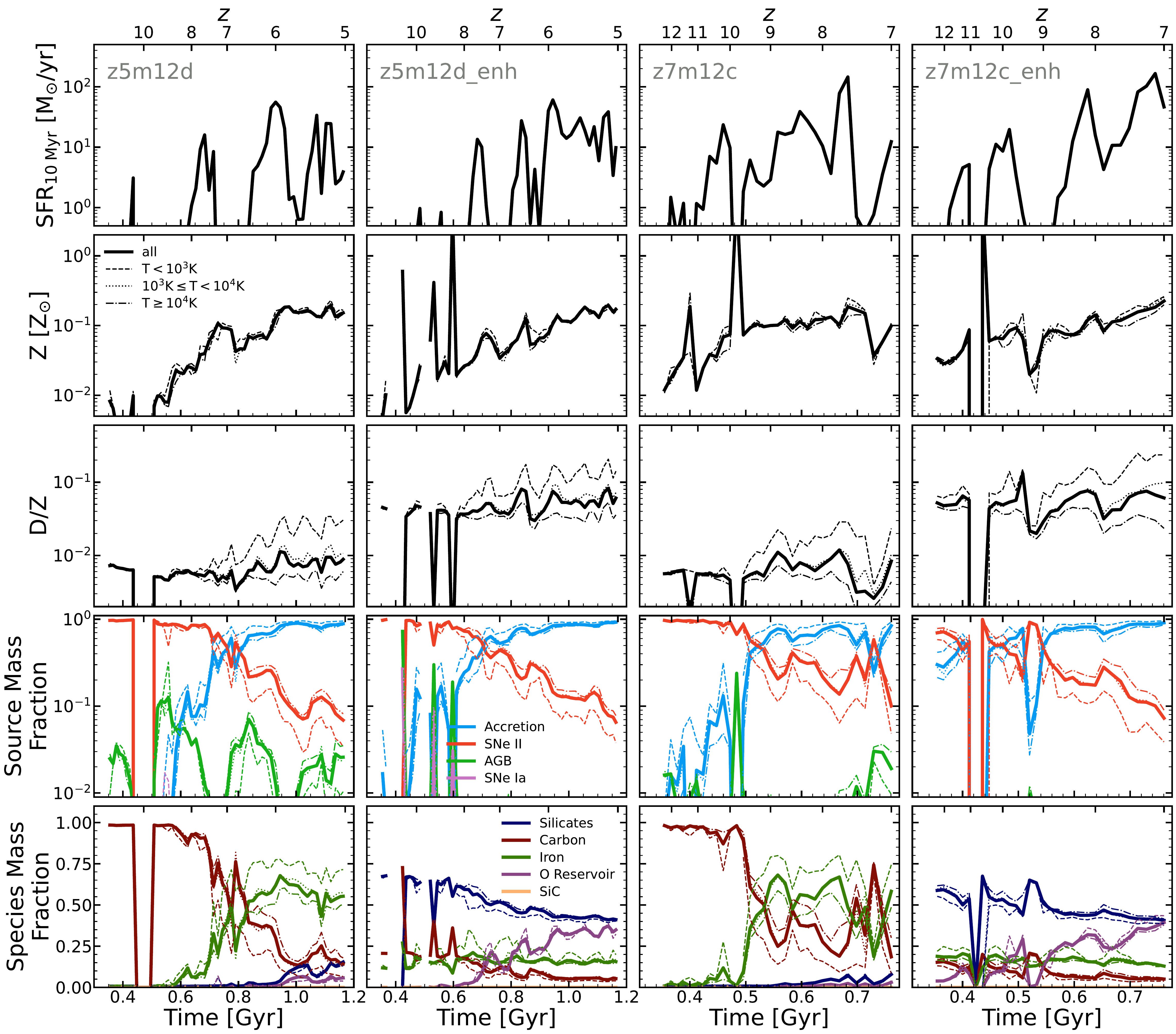}{0.9}
    \vspace{-0.25cm}
    \caption{Same as Fig.~\ref{fig:enhanced_dust_evolution} for {\bf z5m12d} and {\bf z7m12c}.}
    \label{fig:app_dust_evo_comparison2}
\end{figure*}

\begin{figure*}
    \centering
    \plotsidesize{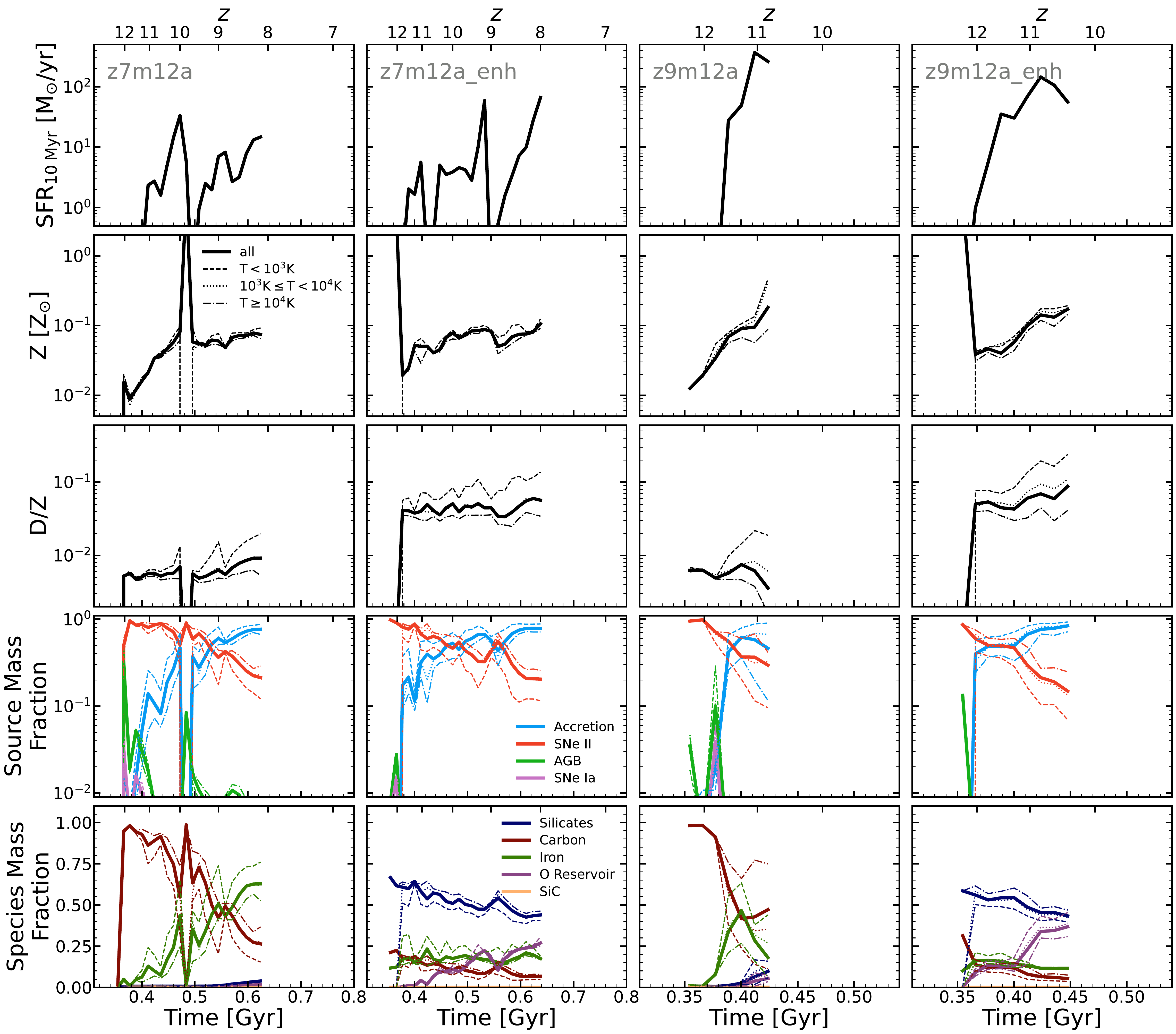}{0.9}
    \vspace{-0.25cm}
    \caption{Same as Fig.~\ref{fig:enhanced_dust_evolution} for {\bf z7m12a} and {\bf z9m12a}.}
    \label{fig:app_dust_evo_comparison3}
\end{figure*}

This appendix shows the detailed evolution of each galaxy's metal and dust population, comparing results from the fiducial and enhanced dust evolution model similar to Fig.~\ref{fig:enhanced_dust_evolution} for the rest of the galaxies in our simulation suite. Fig.~\ref{fig:app_dust_evo_comparison1}, ~\ref{fig:app_dust_evo_comparison2}, and ~\ref{fig:app_dust_evo_comparison3} show the evolution for {\bf z5m11d} and {\bf z5m12d}, {\bf z5m12a} and {\bf z7m12b}, and {\bf z7m12a} and {\bf z9m12a} respectively.

\bsp	
\label{lastpage}
\end{document}